\documentclass[12pt]{article}

\usepackage{amsmath}
\usepackage{indentfirst}
\usepackage{amsfonts,amssymb}
\usepackage{mathrsfs}
\usepackage{CJK,CJKnumb}
\usepackage{color}
\usepackage{amsthm}
\usepackage{graphicx}
\usepackage{enumerate}

\numberwithin{equation}{section}

\def\sF{{\mathscr F}}

\def\sQ{{\mathscr Q}}
\def\sM{{\mathscr M}}

\def\sB{{\mathscr B}}
\def\sX{{\mathscr X}}
\def\sY{{\mathscr Y}}

\def\sC{{\mathscr C}}

\begin{document}

\title{ {On conditional distortion risk measures
                                         \\ under uncertainty}
       \thanks{ Supported in part by the National Natural Science Foundation of China (Nos: 12271415, 12001411)
                   and the Fundamental Research Funds for the Central Universities of China (WUT: 2021IVB024).}\\
       \footnotetext{ Email addresses: shuogong@whu.edu.cn (S. Gong), yjhu.math@whu.edu.cn (Y. Hu), lxwei@whut.edu.cn (L. Wei).  \\
                      \indent\ $^\ast$Corresponding author: Linxiao Wei.
                     }
      }

\vspace{0.5cm}

\author{Shuo Gong$^{1}$, \quad Yijun Hu$^{1}$, \quad Linxiao Wei$^{2, \ast}$ \\
\\
$^1$ {School of Mathematics and Statistics}\\
{Wuhan University} \\
{Wuhan, Hubei 430072}\\
{People's Republic of China}\\
\\
$^2$ {School of Mathematics and Statistics}\\
{Wuhan University of Technology} \\
{Wuhan, Hubei 430070}\\
{People's Republic of China}
}

\date{\today}

\maketitle

\vspace{0.1cm}

\noindent{\bf Abstract:} \quad
Model uncertainty has been one prominent issue both in the theory of risk measures and in practice such as financial risk management
and regulation. Motivated by this observation, in this paper, we take a new perspective to describe the model uncertainty,
and thus propose a new class of risk measures under model uncertainty. More precisely, we use an auxiliary random variable to describe
the model uncertainty.
We first establish a conditional distortion risk measure under an auxiliary random variable. Then we axiomatically characterize it
by proposing a set of new axioms.
Moreover, its coherence and dual representation are investigated.
Finally, we make comparisons with some known risk measures such as weighted value at risk (WVaR), range value at risk (RVaR) and $\sQ-$mixture
of ES. One advantage of our modeling is in its flexibility, as the auxiliary random variable can describe various contexts including
model uncertainty.
To illustrate the proposed framework, we also deduce new risk measures in the presence of background risk.
This paper provides some theoretical results about risk measures under model uncertainty, being expected to make meaningful complement
to the study of risk measures under model uncertainty.

\vspace{0.1cm}


\vspace{0.1cm}


\newpage

\section{Introduction}\label{sec:1}

Risk measures are widely used in various contexts in both finance and insurance, such as regulatory capital calculation and insurance pricing, etc.
Classical risk measures are defined on univariate risks, i.e. on random variables defined on some measurable space $(\Omega, \sF)$,
for instance, see Artzner et al. (1999), F\"{o}llmer and Schied (2002), Frittelli and Rosazza Gianin (2002) and Wang et al. (1997).
In practice, these classical risk measures usually require the acquirement of accurate distribution functions of the random variables.
Mathematically, provided with the random variables, it equivalently requires an accurate probability measure
on $(\Omega, \sF),$ which is sometimes known as a reference probability or a scenario in the literature.
For instance, this is the case of value-at-risk (VaR) or an Expected Shortfall (ES), which are standard risk
measures popularly used in finance and insurance. From the practical perspective, it is usually difficult to accurately capture the \textit{true}
distribution functions of the random variables, because the distribution functions usually have to be estimated from dada or statistical simulation. Alternatively,
it is usually inadequate to theoretically assume that the distribution functions of random variables are accurately known.
Therefore, model uncertainty problem naturally arises.
For more details about risk measures, we refer to F\"{o}llmer and Schied (2016).

\vspace{0.2cm}

Over the past decade-and-a-half, model uncertainty has been attracting more and more attention. Wang and Ziegel (2021) studied scenario-based
risk evaluation, by introducing
\textit{ $\sQ-$based risk measures, $\underline{Q}-$distortion risk measures and $\sQ-$mixture of ES, }
in which $\sQ$ is a collection of finitely many probability measures (secnorios) on some measurable space. Kou and Peng (2016) dealt with
model uncertainty by considering multiple models (scenarios) and introducing \textit{scenario aggregation function.}
Recently, Fadina et al. (2023) further proposes a unified axiomatic framework for generalized risk measures which quantifies jointly a loss random
variable and a set of plausible probabilities (scenarios), in which the set of plausible probabilities describes model uncertainty.
For more earlier studies on model uncertainty problem, we refer to Gilboa and Schmeidler (1989), Hansen and Sargent (2001, 2007),
Zhu and Fukushima (2009), Zymler et al. (2012), Adrian and Brunnermeier (2016) and the references therein.

\vspace{0.2cm}

In the present paper, we take a different perspective in modeling model uncertainty.
Unlike the way of describing model uncertainty taken by Wang and Ziegel (2021) and Kou and Peng (2016), instead of directly assuming the existence
of several probability measures (scenarios) on some measurable space, we use an auxiliary random variable to describe
model uncertainty. For example, each value the auxiliary random variable takes responds a \textit{scenario} on some measurable space.
Our approach is to respectively evaluate the risk of a loss random variable under the condition that the auxiliary random variable takes different
values first, and then aggregate those risk evaluations into a single value.
Although aggregation procedures were also employed in Wang and Ziegel (2021) and Kou and Peng (2016), the aggregation procedure in this paper is totally
different from theirs. One advantage of our modeling is in its flexibility. The auxiliary random variable can describe not only different scenarios
(see Example 4.2 below), but also other risk factors. For instance, the auxiliary random variable can describe the background risk or
a financial institution's risk preference, see Subsection 3.3 and Example 4.1 below. From the mathematical perspective, the auxiliary random variable
can also describe other economical or financial risk factors, say interest rate or the rate of exchange, etc.
Since such an auxiliary random variable serves like a role of \textit{condition} or \textit{environment}, we also call it a
\textit{random environment}, and the resulting risk measures the \textit{conditional risk measures under random environment} in the sequel.

\vspace{0.2cm}

The aim of this paper is to propose a novel axiomatic approach to evaluating risk of loss random variables provided with a random environment.
To our best knowledge, there are no relevant reports available in the literature, and thus we believe that it is worth studying.

\vspace{0.2cm}

The class of distortion risk measures is one of the most important classes of classical risk measures, since it contains a rich family of risk measures
including common VaR and ES, for example, see Belles-Sampera et al. (2014), where term tail value-at-risk (TVaR) was used
instead of ES, F\"{o}llmer and Schied (2016), or Wang and Ziegel (2021).
Particularly, when the distortion functions are concave,
then the distortion risk measures are coherent in the sense of Artzner et al. (1999), for instance, see Dennegerg (1994), Wang et al. (1997)
or  F\"{o}llmer and Schied (2016). From both the theoretical and practical point of view, it is quite natural and reasonable
to further study distortion risk measures under model uncertainty.
Therefore, this paper strives to present conditional distortion-type risk measures under random environment.

\vspace{0.2cm}

In the present paper, we first construct a conditional distortion risk measure under a random environment, which is expressed in terms of
a repeated Choquet integral. Then after its fundamental properties are discussed, we axiomatically characterize it by proposing a set of new axioms,
see Axioms  B1-B4 below.
Although the new axioms have the same flavor as that of classical ones, the new axioms are presented in the presence of a random environment.
Moreover, its coherence and dual representation are investigated. To illustrate the proposed framework for risk measures under model uncertainty,
we also deduce new risk measures in the presence of background risk, see Subsection 3.3 below.
Finally, we make comparisons with some known risk measures such as weighted value at risk (WVaR), range value at risk (RVaR) and $\sQ-$mixture of ES,
see Section 4 below.

\vspace{0.2cm}

We would like to mention that besides the aforementioned scenario-based risk measures, the issue of model uncertainty may have different disguises
in different situations. For instance, Tsanakas (2008) introduced risk measures in the presence of a background risk, where the background risk
could also be understood to describe model uncertainty. Acharya et al. (2017) introduced various systemic risk measures by means of
conditional expectation to evaluate systemic risk, where various conditional probabilities could be supposed to describe model uncertainty.
Recently, Geng et al. (2024) introduces and studies various VaR- and expectile-based systemic risk measures, by providing a unified asymptotic
treatment for systemic risk measures, where various conditional probabilities could also be understood to describe model uncertainty.
Assa and Liu (2024) introduces and characterizes factor risk measures, in which (random) factors could be used to describe the uncertainty of the model.
It is also worth mentioning that our characterization based on state-wise bases is very different from those of Assa and Liu (2024),
where the expressions are based on the distortion functionals on a set of Borel measurable functions.

\vspace{0.2cm}

It should also be mentioned that conditional risk measures have been studied to evaluate systemic risk in the literature, where various conditional
probabilities could be understood to describe model uncertainty.
These conditional risk measures usually involve two random variables: one representing systemic risk (i.e. the overall financial system risk)
and the other representing an individual risk (i.e. a financial institution's risk), and usually focus on the investigation about
the impact of an individual risk on systemic risk and vice versa.
For instance, recently, Dhaene et al. (2022) established a significant conditional distortion risk measure by means of
conditional distribution function of a random variable conditional on the occurrence of an event determined by another random variable; see
Dhaene et al. (2022, Definition 3.1 and Remark 3.2). Various conditional risk measures also appear in Mainik and Schaanning (2014),
Adrian and Brunnermeier (2016), Kleinow et al. (2017) in different disguises, just name a few.
In the above literature, only conditional probability or conditional distribution function is needed.
In contrast to above literature, our approach needs regular conditional probability to ensure relevant measurability issue so that
a Choquet-integral-based aggregation with respect to a random environment can be exercised.
Moreover, by virtue of regular conditional probability, we also provide axiomatic characterization for the introduced conditional distortion
risk measures under model uncertainty.
Taking above consideration into account, the present study can also be viewed as a meaningful complement to the study of conditional risk measures.

\vspace{0.2cm}

It might be helpful to briefly comment on the main contributions of this paper.
First, from a different angle than that of Wang and Ziegel (2021) and Kou and Peng (2016), we propose a new approach to describing model uncertainty
in the course of risk evaluation. Although the present study is mainly motivated by Wang and Ziegel (2021) and Kou and Peng (2016), the way of
describing model uncertainty presented in this paper is different from theirs. One advantage of our approach is in its flexibility,
because the auxiliary random variable (random environment) can describe various contexts including different probability measures (scenarios);
see Subsection 3.3, Examples 4.1.and 4.2 below.
Second, new axioms are presented in the presence of a random environment, which significantly generalize corresponding classical ones. Compared with
the classical ones, these new axioms are more delicate; see Axioms B1-B4 below. With the help of these new axioms, we characterize
a newly introduced class of conditional distortion risk measures, which includes some popular risk measures as special cases,
such as WVaR, RVaR and $\sQ-$mixture of ES, under the help of a plausible condition; see Examples 4.1 and 4.2 below.
Third, new arguments are developed to show the main results of this paper. Since now there is a random environment, new arguments need to be developed
to show the main results. For instance, inspired by Denneberg (1994), we develop a new approach to
showing the existence of the desired monotone set function. Similarly, inspired by Wang et al. (1997) (or Wang and Ziegel (2021)), we develop a new
approach to showing the existence of the desired distortion function.  Compared with those of Denneberg (1994) and Wang et al. (1997)
(or Wang and Ziegel (2021)), these newly developed arguments are far more delicate and complicated due to the presence of the random environment;
see the proofs of Theorems 3.1 and 3.2 below. Thus, the present new arguments can also be viewed as a meaningful development to that of Denneberg (1994)
and Wang et al. (1997).

\vspace{0.2cm}

The rest of this paper is organized as follows. In Section 2, we prepare preliminaries including
definitions, notations and a measurability lemma. Section 3 is devoted to the main results of this paper. As an application, distortion risk measures
in the presence of background risk are also introduced. In Section 4, we make comparisons with some known risk measures such as WVaR, RVaR and
$\sQ-$mixture of ES. Concluding remarks are summarized in Section 5.  In the appendix, we provide the proofs of
all main results presented in Section 3.

\section{Preliminaries}

Let $(\Omega, \sF)$ be a measurable space, and $P$ a fixed probability measure on it, acting as a reference measure,
which is also known as a scenario, for instance, see Wang and Ziegel (2021, Section 3) and Kou and Peng (2016, Section 3).
We denote by $\sX$ the linear space of all bounded measurable functions (i.e. random variables) on $(\Omega, \sF)$
equipped with the supremum norm $\|\cdot\|$, and by $\sX_+$ the subset of $\sX$ consisting of those elements which are non-negative.
In order to deal with model uncertainty, we work with $\sX$ rather than $L^\infty(\Omega, \sF, P)$
of essentially bounded random variables on $(\Omega, \sF, P)$,
for instance, see F\"{o}llmer and Schied (2016, Chapter 4), Wang and Ziegel (2021, Section 3) and Kou and Peng (2016, Section 3).
Throughout this paper, we assume that $(\Omega, \sF, P)$ is an atomless probability space.
The random loss of a financial position is described by an element in $\sX$. For any $X \in \sX$, we denote by $\mbox{Ran}(X)$ the range of $X$,
by $P_X := P \circ X^{-1}$ the probability distribution of $X$ with respect to $P$,
that is, $P_X(A) := P \circ X^{-1} (A) := P(X\in A)$ for any $A \in \sB(\mathbb{R})$, the Borel algebra of subsets of the real line $\mathbb{R}$,
and by $F_X(x) := P(X \leq x), x \in \mathbb{R},$ the distribution function of $X$ with respect to $P$.
For an integrable random variable $X$ on $(\Omega, \sF, P)$, a random variable $Z \in \sX$ and each $z \in \mbox{Ran}(Z)$,
we denote by $E[X|Z]$ and $E[X|Z=z]$ the conditional expectations of $X$ with respect to $Z$ and the event $\{Z=z\}$ under $P$, respectively.
For $A \in \sF,$ $X, X_n \in \sX, n \geq 1,$  we say that the sequence $\{ X_n; n\geq 1\}$ increases to $X$ on $A$, denoted by $X_n \uparrow X$
on $A$, if for all $n \geq 1$ and every $\omega \in A,$ $X_n(\omega) \leq X_{n+1}(\omega)$
and $ \underset{n \rightarrow +\infty}\lim X_n(\omega) = X(\omega).$
For $X, X_n \in \sX, n \geq 1,$  we call that the sequence $\{ X_n; n\geq 1\}$ eventually increases to $X$, denoted by $X_n \uparrow X$ eventually,
if for every $\omega \in \Omega,$ there is an integer  $N := N(\omega) \geq 1,$ so that for all $n \geq N,$ $X_n(\omega) \leq X_{n+1}(\omega)$ and
$ \underset{n \rightarrow +\infty}\lim X_n(\omega) = X(\omega).$
By a U$[0, 1]$ random variable on $(\Omega, \sF, P)$ we mean a random variable which is uniformly distributed on $[0, 1].$
Define the set
\begin{align*}
\sX^\perp  := \{X \in \sX : \ & \mbox{there\ exists\ a\ U} [0, 1] \ \mbox{random\ variable} \\
                               & \mbox{on} \ (\Omega, \sF, P)\ \mbox{independent\ of}\ X\}.
\end{align*}
This set will serve as the random environments. Note that any discrete random variable $X \in \sX$ belongs to $\sX^\perp ;$ for example,
see Lemma 3 of Liu et at. (2020). Note also that $\sX^\perp $ may be a subset of $\sX$ which does not coincide with $\sX.$ For the case
allowing for $\sX^\perp  = \sX$, we refer to Example 9 of Liu et at. (2020).

\vspace{0.2cm}

We introduce more notations. For $a,b \in \mathbb{R}$, $a \vee b$ stands for $\max\{ a,b \}$, and $a \wedge b $ means $ \min \{ a,b \}$.
For $X, Y \in \sX$, $X \vee Y $ stands for $ \max\{ X, Y \},$ and $X \wedge Y $ means $ \min\{ X, Y \}.$ For $Z \in \sX^\perp ,$ by a
$P_Z-$null set $N \in \sB(\mathbb{R})$ we mean that $P_Z(N) := P(Z \in N) = 0.$
Let $D \subseteq \mathbb{R}$ be a non-empty set, denote by $\sB(D)$ the Borel algebra of Borel subsets of $D$, that is,
$\sB(D) := \sB(\mathbb{R})\cap D := \{B\cap D: B \in \sB(\mathbb{R})\}.$ Thus $(D, \sB(D))$ is a measurable space.
For a set $A$, $A^c$ means the complement set of $A,$ and $1_A$ stands for the indicator function of $A$, while $1_{\emptyset} := 0$
with convention. $\mathbb{R}_+ := [0, +\infty).$

\vspace{0.2cm}

Throughout this paper, we assume that for any random variable $Z \in \sX$, the regular conditional probability
$\{ p(z,\cdot) : z\in \mathbb{R} \}$ with respect to $Z$ exists, that is,
for each $z \in \mathbb{R}$, $p(z,\cdot)$ is a probability measure on $(\Omega, \sF)$;
for each $A \in \sF$, $p(\cdot, A)$ is a Borel function on $(\mathbb{R}, \sB(\mathbb{R}))$,
and for any integrable random variable $X$ on $(\Omega, \sF, P)$,
\begin{align}\label{0415add1}
    E[X|Z=z] = \int_{\Omega} X(\omega) p(z,d\omega) \quad \mbox{for} \ P_Z-a.e. \ z \in \mathbb{R}.
\end{align}
Notice that for a general measurable space $(\Omega, \sF)$, there might not exist a system $\{ p(z,\cdot) : z\in \mathbb{R} \}$
which ensures that (\ref{0415add1}) holds. Therefore, we need to technically assume the existence of such a regular conditional probability.
Nevertheless, there are sufficient conditions that ensure the existence of such a regular conditional probability. For instance,
if $(\Omega, \sF)$ is a standard measurable space, then for any random variable $Z$ on $(\Omega, \sF, P)$,
the regular conditional probability $\{ p(z,\cdot) :  z\in \mathbb{R} \}$ with respect to $Z$ exists. Moreover,
a Polish space (i.e. a complete separable metric space) with the topological $\sigma$-algebra is a standard measurable space;
for more details, we refer to Ikeda and Watanabe (1981, pages 11-15).

\vspace{0.2cm}

Any mapping  $\mu$ : $ \sF \rightarrow \mathbb{R}_+$  with $\mu (\emptyset) = 0$ is called a set function on $\sF$.
A set function  $\mu$ on $\sF$ is called normalized, if $\mu(\Omega) = 1,$ and  is called monotone,
if $\mu (A) \leq \mu (B)$ for any $A, B \in \sF$ with $ A \subseteq B$.
Given a monotone set function $\mu$ on $\sF$, for any $X \in \sX$, the Choquet integral of $X$ with respect to $\mu$ is defined as
\begin{align*}
  \int Xd\mu := \int X(\omega)\mu(d\omega) := \int_{-\infty}^0 [\mu(X>x) - \mu(\Omega)] dx + \int_0^{+\infty} \mu(X>x) dx.
\end{align*}
When $\mu$ is taken as a distorted probability $g\circ P$, where $g: [0,1]\rightarrow [0,1]$ is a non-decreasing function with
$g(0)=0$ and $ g(1)=1$, the corresponding
Choquet integral with respect to $g\circ P$ is known as a distortion risk measure, and $g$ is called a distortion function. For example,
see Yaari (1987), Denneberg (1990), Wang (1996), Wang et al. (1997), Acerbi (2002), Belles-Sampera et al. (2014) and F\"{o}llmer and Schied (2016).
Note also that the requirement of monotonicity of a distortion function is not necessary in general, for instance, see Wang et al. (2020).

\vspace{0.2cm}

Given an $\alpha \in (0,1)$, for any $ X \in \sX$, the VaR of $X$ at the confidence or tolerance level $\alpha$ is defined
by
\begin{align*}
  \mbox{VaR}_{\alpha}(X) &:= \inf\{x : F_X(x)\geq \alpha\},
\end{align*}
and the ES of $X$ at the confidence or tolerance level $\alpha$ is defined by
\begin{align*}
\mbox{ES}_{\alpha}(X) &:= \frac{1}{1-\alpha} \int_{\alpha}^1 \mbox{VaR}_{\theta}(X) d\theta.
\end{align*}
We also use the notation $\mbox{ES}_{\alpha}^P(X)$ to emphasize the being used probability measure (scenario) $P$; see Subsection 4.2 below.

\vspace{0.2cm}

We now introduce a notion of regularity for a family of mappings, which is needed later.

\vspace{0.2cm}

\noindent{\bf Definition 2.1}\ \  
Let $D \subseteq \mathbb{R}$ be a non-empty set, and
$\{ f_d ; d \in D \}$ be a family of mappings from a common domain $\sY$ to $\mathbb{R}$.
The family $\{ f_d ; d \in D \}$ is called regular on $D$, if for any fixed $y \in \sY$, the function $d \rightarrow f_d(y)$
is $\sB(D)$-measurable, that is, it is a Borel function on $D$.

\vspace{0.2cm}

Comonotonicity is an important notion in finance and insurance literature. We introduce a notion of local comonotonicity for random variables.

\vspace{0.2cm}

\noindent{\bf Definition 2.2} \ \ Let $A \in \sF.$ For $X, Y \in \sX$, $X$ and $Y$ are called local-comonotonic on $A$,
if for every $\omega_1,\omega_2 \in A$,
\begin{align*}
  (X(\omega_1)-X(\omega_2))(Y(\omega_1)-Y(\omega_2)) \geq 0.
\end{align*}
Particularly, we simply say that $X$ and $Y$ are comonotonic on $\Omega$, if they are local-comonotonic on $\Omega$.

\vspace{0.2cm}

We now introduce the definition of conditional risk measures under random environment.

\vspace{0.2cm}

\noindent{\bf Definition 2.3 }\ \
A conditional risk measure under random environment is defined as any functional $\rho (X; Z)$ : $\sX \times \sX^\perp  \rightarrow \mathbb{R}$.
Particularly, for any random loss $X \in \sX$ and any random environment $Z \in \sX^\perp $,
the quantity $\rho (X; Z)$ is called the risk measure of  $X$ under random environment $Z$.

\vspace{0.2cm}

Notice that in the above definition of conditional risk measures under random environment, the two input arguments take different roles.
More precisely, the first input argument represents the random loss of a financial position. Relative to the first input argument,
the second input argument serves only as a role of sort of \textit{condition} or \textit{environment}.
Different elements in $ \sX^\perp $ represent different  \textit{conditions (environments)}.
In the majority of cases,  we specify a \textit{condition (environment)} $Z \in \sX^\perp $ first,
and then evaluate the risk of random losses $X \in \sX$.
In other words, what we do is to evaluate the risk of random losses $X$ provided with a \textit{condition} (\textit{environment})
$Z \in \sX^\perp ;$ for instance, see Subsection 3.2 below for the discussion about coherence and dual presentation,
where an alternative notation $\rho_{_Z}(X)$ is used,
and $\rho_{_Z}(\cdot)$ is considered as a \textit{univariate} functional of random loss; see also Example 4.2 below for the comparison with
$\sQ$-mixture of ES, where a \textit{condition} (\textit{environment}) is specified first according to the given probability measures (scenarios),
and then we evaluate the risk of random losses provided with the \textit{condition} (\textit{environment}).
Since the \textit{condition (environment)} $Z \in \sX^\perp $ is a \textit{random} variable,
this is also the reason why we refer to the second input argument $Z$ as \textit{random environment}, and to the risk measure $\rho$
as \textit{conditional} risk measure under \textit{random environment}. Nevertheless, considering that the random environment $Z$ can be any element
in $\sX^\perp ,$ at this moment, we use notation $\rho (X; Z)$ rather than others, and regard $\rho$ as a \textit{bivariate} functional of
random loss and random environment. For such a viewpoint of considering $\rho$ as a \textit{bivariate} functional, one mathematical advantage is
in its flexibility. For instance, the specification of a random environment is also allowed to depend on a given random loss;
see Subsection 3.3 below for the risk measures in the presence of \textit{background risk}, where the random environment is specified to be
the sum of the \textit{background risk} and a given random loss; see also Example 4.1 below for the comparisons with WVaR and RVaR,
where the specification of a random environment is related to a given random loss.

\vspace{0.2cm}

For terminology convenience, the values that a random environment $Z$ takes are also called the \textit{states} of $Z$.
For any random environment $ Z \in \sX^\perp  $, denote by $\{ K_{Z}(z,\cdot) : z\in \mathbb{R} \}$ the regular conditional probability
with respect to $Z$.
By the properties of regular conditional probability $\{ K_Z(z, \cdot) : z \in \mathbb{R} \}$,
there exists a $P_Z-$null set $N_0 \in \sB(\mathbb{R})$ such that for every $z \in N_0^c, $
\begin{align}\label{0429add1}
  K_Z(z,\{ Z \in B \}) = 1_{B}(z) \quad \mbox{for\ any} \ B \in \sB(\mathbb{R}),
\end{align}
for instance, see Ikeda and Watanabe (1981, Corollary of Theorem 3.3, page 15).
Therefore, for every $z \in N_0^c,$ $K_Z(z, \{ Z =z\}) = 1.$ Consequently,
for each $z\in N_0^c \cap \mbox{Ran}(Z)$, we denote by $ K_Z^*(z, \cdot)$ the restriction of probability measure $ K_Z(z, \cdot)$
to the $\sigma$-algebra $ \sF \cap \{Z = z\} := \{A \cap \{Z = z\} :\ A \in \sF \}$
such that $(\{Z = z\},  \sF \cap \{Z = z\}, K_Z^*(z, \cdot))$ is a probability space,
where for any $A^* := A \cap \{Z = z \} \in \sF \cap \{Z = z\}$ with some $A \in \sF $, $K_Z^*(z, A^*) := K_Z(z, A)$.
Similarly, we denote by $P_Z^*$ the restriction of $P_Z$ to the Borel algebra $\sB(\mbox{Ran}(Z))$ such that
$(\mbox{Ran}(Z), \sB(\mbox{Ran}(Z)), P_Z^*)$ is a probability space, where for any $B^* := B \cap \mbox{Ran}(Z) \in \sB(\mbox{Ran}(Z))$
with some $B \in \sB(\mathbb{R})$, $P_Z^*(B^*) := P_Z(B)$.
By $P_Z-a.e.$ $z \in \mbox{Ran}(Z)$ we mean that there is a $P_Z-$null set $B_0 \in \sB(\mathbb{R})$ with $N_0 \subseteq B_0$
such that $z \in B_0^c \cap \mbox{Ran}(Z)$, where the $P_Z-$null set $N_0$ is as in ($\ref{0429add1}$).
Similarly, given a state $z \in \mbox{Ran}(Z)$, by $K_Z(z, \cdot)-a.e.$ $\omega \in \{Z= z\}$
we mean that there is an $\Omega_0 \in \sF$ with $K_Z(z, \Omega_0) = 0$ such that $\omega \in \Omega_0^c \cap \{Z= z\}$.
When the random environment $Z$ is degenerate, that is, $Z$ takes sole possible value, say $z_0$, then $K_Z(z_0,\cdot)$ is just
the probability measure $P$, and hence $K_Z(z, \cdot)$ can be chosen to be identical to $P$ for each $z \in \mathbb{R}$.

\vspace{0.2cm}

Now, we turn to introduce the definition of conditional risk measures under fixed state. This notion is a starting point for
our studying conditional risk measures under random environment.

\vspace{0.2cm}

\noindent{\bf Definition 2.4}\ \ Given a random environment $Z \in \sX^\perp  $ and a state
$z \in \mbox{Ran}(Z)$, a conditional risk measure under fixed state $z$ is defined as any functional
$\rho_{_Z}(\cdot; z) : \sX \rightarrow \mathbb{R}$ (or $\sX_+ \rightarrow \mathbb{R}$, respectively).
Particularly, for any  $ X \in \sX$ (or $\sX_+$,  respectively),
the quantity $\rho_{_Z}(X; z)$ is called the risk measure of random loss $X$ under fixed state $z.$

\vspace{0.2cm}

Next, we introduce one more notion of
environment-wise comonotonicity for functions, which is also needed later.

\vspace{0.2cm}

\noindent{\bf Definition 2.5}\ \
Given any random environment $Z \in \sX^\perp $, for any $z \in \mbox{Ran}(Z)$, let $\tau_Z(\cdot;z) : \sX
\rightarrow \mathbb{R}$ be a functional. For any two random losses  $X_1, X_2 \in \sX$, the two functions
$\tau_Z(X_1;\cdot), \tau_Z(X_2;\cdot) : \mbox{Ran}(Z) \rightarrow \mathbb{R}$
are called environment-wise comonotonic, if for every $z_1, z_2 \in \mbox{Ran}(Z)$,
$$
 \left( {\tau_Z}(X_1; z_1)- {\tau_Z}(X_1; z_2) \right) \left( {\tau_Z}(X_2; z_1)- {\tau_Z}(X_2; z_2) \right) \geq 0.
$$

\vspace{0.2cm}

We end this section with a measurability lemma, which is also crucial for us to establish conditional distortion risk measures
under random environment later.
For the purpose of relative self-containedness, we will provide its proof in the appendix, thought the proof itself is not complicated.

\vspace{0.2cm}

\noindent{\bf Lemma 2.1}\ \ Let $g(z,x): \mathbb{R} \times \mathbb{R}_+ \rightarrow \mathbb{R}$ be a function such that for any $x \in \mathbb{R}_+$, $g(\cdot,x)$ is a Borel function on $\mathbb{R},$  and for any $z \in \mathbb{R}$, $g(z,\cdot)$ is left-continuous. Then for any Borel function
$\phi : \mathbb{R} \rightarrow \mathbb{R}_+$, the function $z \rightarrow g(z,\phi(z))$ is a Borel function on $\mathbb{R}$.

\section{Main results}

In this section, we present the main results of this paper. We will begin with introducing axioms and the definition of
a conditional distortion risk measure under random environment. Then, after discussing its properties, we axiomatically characterize it.
Moreover, its coherence and dual representation are investigated.
Finally, as an application, we introduce two new distortion risk measures in the presence of background risk.
All proofs of the results of this section will be postponed to the appendix.

\subsection{Conditional distortion risk measures under random environment }

In this subsection, some axioms for conditional risk measures under random environment are listed first.
Then conditional distortion risk measures under random environment are constructed.
After studying its properties, we axiomatically characterize it as well.

\vspace{0.2cm}

We begin with introducing axioms for a conditional risk measure under fixed state $z$
$\rho_{_Z}(\cdot; z)$ : $\sX$ $\rightarrow$ $ \mathbb{R}$.

\vspace{0.1cm}

\begin{enumerate}
\item[A1] State-wise law invariance: Given a random environment $Z$ and a state $z \in $ Ran($Z$),
      for any $X \in \sX,$ the risk measure $\rho_{_Z}(X; z)$ only depends on the probability distribution
      $K_Z(z, \cdot) \circ X^{-1}$ of $X$ with respect to $K_Z(z, \cdot)$, that is, $\rho_{_Z}(X; z)= \rho_{_Z}(Y; z)$ for any
       $ X, Y \in \sX$ with $K_Z(z, \cdot) \circ X^{-1} = K_Z(z, \cdot) \circ Y^{-1}$.

\item[A2] State-wise monotonicity: Given a random environment $Z$ and a state $z \in $ Ran($Z$),
        for random losses $X, Y \in \sX$, if $X(\omega) \leq Y(\omega)$ for $K_Z(z, \cdot)- a.e.$
         $\omega \in \ \{Z = z \}$, then $\rho_{_Z}(X; z) \leq \rho_{_Z}(Y; z)$.

\item[A3] State-wise comonotonic additivity: Given a random environment $Z$ and a state $z \in $ Ran($Z$),
        for two random losses $X, Y \in \sX$,  if $X$ and $Y$ are local-comonotonic on $\{ Z = z \}$,
        then $\rho_{_Z}(X+Y; z)= \rho_{_Z}(X; z)+ \rho_{_Z}(Y; z)$.

\item[A4] State-wise continuity from below: Given a random environment $Z$ and a state $z \in $ Ran($Z$),
      for random losses  $X, X_n \in \sX,\ n \geq 1$, if $X_n \uparrow X$ on $\{ Z=z \}$, then
      $\underset{n \rightarrow \infty}{\lim} $ $ \rho_{_Z}(X_n; z) $ $ = \rho_{_Z}(X; z)$.
\end{enumerate}

\vspace{0.1cm}

Since the state of a given environment is pre-specified, the financial meanings of above Axioms A1-A4 can be interpreted totally similarly
to the classic setting; for instance, see Artzner et al. (1999), Wang et al. (1997), Kusuoka (2001), F\"{o}llmer and Schied (2016)
and Kou and Peng (2016).

\vspace{0.2cm}

Next, we proceed to introduce axioms for conditional risk measures under random environment
$\rho(\cdot; \cdot) : \sX \times \sX^\perp  \rightarrow \mathbb{R}$. For this purpose, we need one more assumption as follows:

\vspace{0.2cm}

\noindent{\bf Assumption A}\ \ Given any random environment $Z \in \sX^\perp $, for each $z \in \mbox{Ran}(Z)$, there exists a functional
$\tau_Z(\cdot; z) : \sX \rightarrow \mathbb{R}$ with $\tau_Z(1; z) = 1$,
such that the family $\{ \tau_Z(\cdot; z) ; z \in \mbox{Ran}(Z) \}$ of functionals is regular on $\mbox{Ran}(Z).$
Moreover, for $P_Z-a.e.$ $z \in \mbox{Ran}(Z),$ the functional $\tau_Z(\cdot; z)$ satisfies Axioms A1--A4.

\vspace{0.2cm}

Note that the existence of such a family $\{ \tau_Z(\cdot; z) ; z \in \mbox{Ran}(Z) \}$ as in Assumption A will be demonstrated later,
see Definition 3.2(1), Remark 3.2(i) and Proposition 3.1(1) below.

\vspace{0.2cm}

Now, we are ready to state the axioms for conditional risk measures under random environment
 $\rho(\cdot; \cdot) : \sX \times \sX^\perp  \rightarrow \mathbb{R}$.

\vspace{0.1cm}

\begin{enumerate}
 \item[B1] Environment-wise law invariance: Given any random environment $Z \in \sX^\perp $, for each $z\in \mbox{Ran}(Z)$,
       let $\tau_Z(\cdot; z)$ be the functional as in Assumption A. For two random losses $X_1, X_2 \in \sX$,
        if $\tau_Z(X_1;\cdot)$ and $\tau_Z(X_2;\cdot)$ have the same probability distribution with respect to $P_Z^*$, that is,
        $P_Z^* \circ \tau_Z^{-1}(X_1;\cdot) = P_Z^* \circ \tau_Z^{-1}(X_2;\cdot)$, then $\rho(X_1; Z) = \rho(X_2; Z)$.

 \item[B2] Environment-wise monotonicity: Given any random environment $Z \in \sX^\perp $, for each $z\in \mbox{Ran}(Z)$,
        let $\tau_Z(\cdot; z)$ be the functional as in Assumption A.  For two random losses $X_1, X_2 \in \sX$,
        if $\tau_Z(X_1; z)\leq \tau_Z(X_2; z)$ for $P_Z-a.e.\ z\in \mbox{Ran}(Z)$, then $\rho(X_1; Z) \leq \rho(X_2; Z)$.

 \item[B3] Environment-wise comonotonic additivity:  Given any random environment $Z \in \sX^\perp $, for each $z\in \mbox{Ran}(Z)$,
      let $\tau_Z(\cdot; z)$ be the functional as in Assumption A.  For two random losses $X_1, X_2 \in \sX,$ if for $P_Z-a.e.\ z\in \mbox{Ran}(Z)$,
      $X_1$ and $X_2$ are local-comonotonic on $\{Z = z\}$, and the functions $\tau_Z(X_1; \cdot)$ and $\tau_Z(X_2; \cdot)$ are environment-wise comonotonic,
       then $\rho(X_1 + X_2; Z) = \rho(X_1; Z)+ \rho(X_2; Z)$.

 \item[B4] Environment-wise continuity from below: Given any random environment $Z \in \sX^\perp $,  for each $z\in \mbox{Ran}(Z)$,
      let $\tau_Z(\cdot; z)$ be the functional as in Assumption A. For random losses $X, X_n \in \sX$, $n\geq 1,$
      if the sequence $\{ X_n; n \geq 1 \}$ is bounded below by some constant $D \in \mathbb{R}$ (i.e. $D \leq X_n(\omega)$ for each $n\geq 1$
      and every $\omega \in \Omega$), $X_n \uparrow X$ eventually, and $\tau_Z(X_n; z) \leq \tau_Z(X; z)$ for each $n \geq 1$ and
      $P_Z-a.e.\ z \in \mbox{Ran}(Z)$, then $\underset{n \rightarrow +\infty}{\lim} \rho(X_n; Z) = \rho(X; Z)$.
\end{enumerate}

\vspace{0.1cm}

Axioms B1-B4 can be interpreted in the context of finance as follows.
Note first that $\tau_Z(X; z)$ as in Assumption A exactly represents the risk measure of random loss $X$ under the condition that the
environment $Z$ takes the state $z$, as will be seen in the sequel.
Axiom  B1 means that for two random losses $X_1$ and $X_2$, if their state-wise riskinesses $\tau_Z(X_1; \cdot)$ and $\tau_Z(X_1; \cdot)$
have the same distribution under the environment's probability distribution, then their overall riskinesses should be the same.
This characteristic has some similarity to the law invariance in the classic setting.
B2 says that if random loss $X_1$ is less risky than another random loss $X_2$ in almost-all-state-wise sense,
then the overall riskiness of $X_1$ should not exceed that of $X_2$.
B3 means that if two random losses $X_1$ and $X_2$ are comonotonic in almost-all-state-wise sense, and moreover the two risk measures
$\tau_Z(X_1; z)$ and $\tau_Z(X_2; z)$ are also comonotonic with respect to the state variable $z$, then the overall riskiness of $X_1+X_2$
should be the superposition of those of $X_1$ and $X_2$, because spreading risk within comonotonic risks can not reduce the total risk.
As for B4, from mathematical point of view, it more or less belongs to a kind of technical requirement. Nevertheless, it can still be
interpreted in the financial context as follows.
For a sequence of random losses $X_n$, $n \geq 1,$ if for almost all states $z \in \mbox{Ran}(Z)$, the state-wise riskinesses
$\tau_Z(X_n; z)$ of $X_n$, $n\geq 1,$ have a ceiling
that is just the state-wise riskiness $\tau_Z(X; z)$ of another random loss $X$, meanwhile the sequence of random losses $X_n,$  $n\geq 1$,
eventually increases to the random loss $X$ regardless of what state the environment takes, then the sequence of overall riskinesses of $X_n$,
$n\geq 1$, should also converge to that of $X$.

\vspace{0.2cm}

\noindent{\bf Remark 3.1}\ \
(i) In the course of introducing above Axioms A1-A4 and  B1-B4, if the domains $\sX$
and $\sX \times \sX^\perp $ of risk measures are replaced with $\sX_+$ and
$\sX_+ \times \sX^\perp $, respectively, then the counterparts of these axioms can be parallel introduced.
In the sequel, if needed, those counterparts should be in use. Since there should have no risk of confusion, we do not want to repeat
those counterparts almost verbatim.

\vspace{0.1cm}

(ii) Note that both the state-wise monotonicity A2 and state-wise comonotonic additivity A3 of $\rho_{_Z}(\cdot; z)$
imply the positive homogeneity of $\rho_{_Z}(\cdot; z)$, that is, $\rho_{_Z}(cX; z)= c\cdot \rho_{_Z}(X; z)$ for any $c>0$ and any $X \in \sX$.
Similarly,  both the environment-wise monotonicity B2 and environment-wise comonotonic additivity B3 of $\rho $
imply the positive homogeneity of $\rho(\cdot; Z)$, that is, given a random environment $Z$,
$\rho(cX; Z)= c\cdot \rho(X; Z)$ for any $c>0$ and any  $X \in \sX$. For more details,
we refer to Denneberg (1994, Theorem 11.2 and Exercise 11.1).

\vspace{0.2cm}

\noindent{\bf Definition 3.1 (Normalization)}\ \ (1)\ Given a random environment $Z$ and any $z\in \mbox{Ran}(Z)$,  a conditional risk measure
under fixed state $z$ $\rho_{_Z}(\cdot; z)$ : $\sX$  $\rightarrow$ $ \mathbb{R}$
is called normalized, if $\rho_{_Z}(1;z)= 1$.

\vspace{0.1cm}

(2)\  A conditional risk measure under random environment $\rho(\cdot; \cdot)$ : $\sX \times \sX^\perp  \rightarrow \mathbb{R}$
is called normalized, if $\rho(1; Z)= 1$ for any $Z \in \sX^\perp $.

\vspace{0.2cm}

The normalization can be interpreted as follows. For a degenerate random loss $X =1$,
its capital requirement should reasonably be $1$. While in the context of insurance, the normalization is also known as the
so-called \textsl{no unjustified risk loading,} for example, see Wang et al. (1997).

\vspace{0.2cm}

Next, we introduce the definitions of conditional distortion risk measures (CDRMs) under fixed state and random environment.

\vspace{0.2cm}

\noindent{\bf Definition 3.2}\ \
Let $\{ g_z; z \in \mathbb{R} \}$ be a family of left-continuous distortion functions, which is regular on $\mathbb{R}$.

\vspace{0.1cm}

(1)\ Let $Z \in \sX^\perp $ be a random environment. For each $z\in \mbox{Ran}(Z)$, the normalized conditional distortion risk measure
under fixed state $z$
$\rho_{_Z}(\cdot; z)$ : $\sX$  $\rightarrow$ $ \mathbb{R}$ (or $\sX_+$ $\rightarrow$ $ \mathbb{R}$, respectively)
 is defined by
\begin{align}\label{add3001}
   \rho_{_Z}(X; z) :=  \int_{-\infty}^0 \left[ g_z \circ K_Z(z,\{ X > \alpha \})-1 \right] d\alpha
                   + \int_0^{\infty} g_z \circ K_Z(z,\{ X > \alpha \})d\alpha
\end{align}
for $X \in \sX$ (or $\sX_+$, respectively).

\vspace{0.1cm}

(2)\ For each random environment $Z \in \sX^\perp $, let $h_Z$ be a distortion function associated with $Z$. The normalized conditional
distortion risk measure under random environment
$\rho$ : $\sX \times \sX^\perp  \rightarrow \mathbb{R}$ (or $\sX_+ \times \sX^\perp  \rightarrow \mathbb{R}$, respectively)
 is defined by
\begin{align}\label{add3002}
 \rho(X; Z)
    & := \int_{-\infty}^0 \left[ h_Z \circ P_Z \left( \left\{ z: \int_{-\infty}^0 \left( g_z \circ K_Z(z, \{X > \alpha \})- 1\right) d\alpha
                          \right. \right. \right. \nonumber \\
    & \quad + \left. \left. \left. \int_0^{\infty} g_z \circ K_Z(z, \{X > \alpha \})d\alpha >\beta \right\} \right) - 1 \right] d\beta \nonumber \\
    & \quad + \int_0^{\infty} h_Z \circ P_Z \left( \left\{ z: \int_{-\infty}^0 \left( g_z \circ K_Z(z, \{ X > \alpha \})- 1\right) d\alpha
                                           \right. \right. \nonumber\\
    & \quad + \left. \left. \int_0^{\infty} g_z \circ K_Z(z,\{ X > \alpha \})d\alpha >\beta \right\} \right) d\beta
\end{align}
for $(X, Z) \in \sX \times \sX^\perp $ (or $\sX_+ \times \sX^\perp $, respectively).

\vspace{0.2cm}

\noindent{\bf Remark 3.2}\ \ (i)\ For any fixed $X \in \sX$, by Lemma 2.1, we know that
the right hand side of $(\ref{add3001})$ is a Borel function on $\mathbb{R}$ with respect to variable $z \in \mathbb{R}$.
Hence, the family $\{ \rho_{_Z}(\cdot;z) ; z \in \mathbb{R} \}$ of functionals defined by ($\ref{add3001}$) is regular on $\mathbb{R}$, that is,
for any $X \in \sX$, the function $z \rightarrow \rho_{_Z}(X; z)$ is a Borel function on $\mathbb{R}.$
Thus, the family $\{ \rho_{_Z}(\cdot;z) ; z \in \mbox{Ran}(Z) \}$ is regular on $\mbox{Ran}(Z)$ as well.
Note also that the family $\{ \rho_{_Z}(\cdot; z) ; z \in \mbox{Ran}(Z) \}$ is also regular on $\mbox{Ran}(Z),$  upon the subset
$\{ g_z; z \in \mbox{Ran}(Z) \}$ of $\{ g_z; z \in \mathbb{R} \}$ is regular on $\mbox{Ran}(Z)$.
Furthermore, when $z \in \mbox{Ran}(Z)$, then ($\ref{add3001}$) defines a conditional risk measure under fixed state $z$
in the sense of Definition 2.4, which is transparently normalized, and it is the case what we will really concern in the sequel.

\vspace{0.1cm}

(ii) Taking (\ref{add3001}) into account, we can rewrite $\rho(X; Z)$ defined by ($\ref{add3002}$) in terms of $\rho_{_Z}(\cdot; z)$
as follows:
\begin{align}\label{add3012}
  \rho(X; Z)
    & = \int_{-\infty}^0 \left[ h_Z \circ P_Z \left( \left\{ z: \rho_{_Z}(X;z)>\beta \right\} \right) -1 \right] d\beta
                   + \int_0^{\infty} h_Z \circ P_Z \left( \left\{ z: \rho_{_Z}(X;z)>\beta \right\} \right) d\beta  \nonumber\\
    & = \int \rho_{_Z}(X; \cdot) dh_Z \circ P_Z^*,
\end{align}
for $(X,Z) \in \sX \times \sX^\perp $.

\vspace{0.2cm}

Now, we turn to study the properties of CDRMs under fixed state and random environment $\rho_{_Z}(\cdot; z)$ and $\rho$ defined by (\ref{add3001})
and ($\ref{add3002}$) respectively.

\vspace{0.2cm}

\noindent{\bf Proposition 3.1}\ \  Let $\{g_z ; z \in \mathbb{R} \}$ be a family of left-continuous distortion functions,
which is regular on $\mathbb{R}$.

(1)\ Let $Z \in \sX^\perp $ be an arbitrarily given random environment, and $N_0 \in \sB(\mathbb{R})$ the $P_Z-$null set as in ($\ref{0429add1}$).
For each $z\in \mbox{Ran}(Z)$, let the functional $\rho_{_Z}(\cdot; z)$ be the normalized CDRM under fixed state $z$
defined by $(\ref{add3001})$. Then for every $z\in N_0^c \cap \mbox{Ran}(Z)$,  $\rho_{_Z}(\cdot; z)$  satisfies Axioms A1-A4.

\vspace{0.1cm}

(2)\ For each random environment $Z \in \sX^\perp $, let $h_Z$ be a distortion function associated with $Z$.
For each $z \in \mbox{Ran}(Z)$, let the functional $\tau_Z(\cdot; z) : \sX \rightarrow \mathbb{R}$
in Assumption A be  $\rho_{_Z}(\cdot; z)$ defined by $(\ref{add3001})$.  Then the normalized CDRM under random environment
$\rho$ defined by $(\ref{add3002})$  satisfies Axioms B1-B3.
In addition, if the distortion function $h_Z$ is left-continuous, and each distortion function in the subset $\{ g_z ; z\in \mbox{Ran}(Z) \}$
of $\{ g_z ; z\in \mathbb{R} \}$ is continuous, then $\rho$ further satisfies Axiom B4.

\vspace{0.2cm}

Proposition 3.1(1) says that for $P_Z-a.e.$ $z \in \mbox{Ran}(Z)$, the normalized CDRM under fixed state $z$ $\rho_{_Z}(\cdot; z)$
defined by $(\ref{add3001})$ satisfies Axioms A1-A4, if the distortion function $g_z$ is left-continuous.
The following Propositions 3.2-3.4 give the axiomatic characterization of normalized CDRMs under fixed state,
and they are also known as representation for functionals in the risk measure literature.
Basically, the approaches to these representations are the same as ones in the classic setting;
for instance, see Denneberg (1994), Wang et al. (1997), F\"{o}llmer and Schied (2016) and Kou and Peng (2016).
Nevertheless, since these representations will be crucial to the subsequent study, and play important roles in the study of representations
for CDRMs under random environment, we think it would be necessary and helpful to clarify them in the name of propositions.

\vspace{0.2cm}

\noindent{\bf Proposition 3.2}\ \
Given a random environment $Z\in \sX^\perp $, assume that for each state $z \in \mbox{Ran}(Z)$, there is a normalized conditional risk measure
under fixed state $z$ $\rho_{_Z}(\cdot; z) : \sX \rightarrow \mathbb{R}$.
If for $P_Z-a.e.$ $z \in \mbox{Ran}(Z)$,  $\rho_{_Z}(\cdot; z)$ satisfies Axioms A2 and A3,
then for every such $P_Z-a.e.$ $z \in \mbox{Ran}(Z)$,
there exists a monotone and normalized set function $\gamma_z$ on $\sF$ depending on $z$ and uniquely determined by $\rho_{_Z}(\cdot;z)$
such that for any $X \in \sX_+$,
$$
\rho_{_Z}(X; z) = \int_0^{\infty} \gamma_z(\{ X > \alpha \})d\alpha.
$$
Particularly, for every $A \in \sF,$ $\gamma_z(A) := \rho_{_Z}(1_A; z).$

\vspace{0.2cm}

\noindent{\bf Proposition 3.3}\ \
Given a random environment $Z\in \sX^\perp $, assume that for each state $z \in \mbox{Ran}(Z)$, there is a normalized conditional risk measure
under fixed state $z$ $\rho_{_Z}(\cdot; z) : \sX \rightarrow \mathbb{R}$.
If for $P_Z-a.e.$ $z \in \mbox{Ran}(Z)$,  $\rho_{_Z}(\cdot; z)$  satisfies Axioms A1-A4,
then there is a $P_Z-$null set $N \in \sB(\mathbb{R})$, so that for every $z \in N^c \cap \mbox{Ran}(Z)$, there exists a left-continuous
distortion function $g_z$ depending on $z$ such that for any $X \in \sX_+$,
$$
\rho_{_Z}(X; z) = \int_0^{\infty} g_z \circ K_Z(z,\{ X > \alpha \})d\alpha.
$$
Particularly, for $u \in [0,1],$ $ g_z(u) := \rho_{_Z} \left( 1_{\{ U > 1-u \}}; z \right),$ where $U$ is a U$[0,1]$ random variable
on $(\Omega, \sF, P)$ independent of $Z$.

\vspace{0.2cm}

Next proposition extends Proposition 3.3 to general random losses.

\vspace{0.2cm}

\noindent{\bf Proposition 3.4}\ \
Given a random environment $Z\in \sX^\perp $, assume that for each state $z \in \mbox{Ran}(Z)$, there is a normalized conditional risk measure
under fixed state $z$ $\rho_{_Z}(\cdot; z) : \sX \rightarrow \mathbb{R}$.
If for $P_Z-a.e.$ $z \in \mbox{Ran}(Z)$, $\rho_{_Z}(\cdot; z)$ satisfies Axioms A1-A4,
then there is a $P_Z-$null set $N \in \sB(\mathbb{R})$ as in Proposition 3.3, so that for every $z \in N^c \cap \mbox{Ran}(Z)$,
there exists a left-continuous distortion function $g_z$ depending on $z$
such that for any $ X \in \sX$,
\begin{align*}
\rho_{_Z}( X; z)
         = \int_{-\infty}^0 (g_z \circ K_Z(z,\{ X > \alpha \})-1)d\alpha
             + \int_0^{\infty} g_z \circ K_Z(z,\{ X > \alpha \})d\alpha,
\end{align*}
where the distortion function $g_z$ is given as in Proposition 3.3.

\vspace{0.2cm}

\noindent{\bf Remark 3.3}\ \ As pointed out in the previous section, when the random environment $Z$ is degenerate, then for each $z \in \mathbb{R}$, $K_Z(z,\cdot) = P(\cdot)$.
Intuitively, in this case, the \textit{randomness} of the environment disappears, and thus Propositions 3.3 and 3.4 coincide with
the classic ones; for instance, see Wang et al. (1997).
The main distinction between this paper and that of Wang et al. (1997) lies in the assumptions employed.
More precisely, given a random environment $Z \in \sX^\perp $, $\sX_+$ contains a continuous random variable under probability measure $K_Z(z, \cdot)$
for $P_Z-a.e.$ $z\in \mbox{Ran}(Z)$, whereas Wang et al. (1997) assumed that the collection of risks contains all the Bernoulli($u$) random variables,
$0\leq u \leq 1.$

\vspace{0.2cm}

Proposition 3.1(2) says that the normalized CDRM under random environment $\rho$ defined by $(\ref{add3002})$ satisfies Axioms B1-B4,
upon the distortion function $h_Z$ is left-continuous, and each distortion function in $\{g_z, z\in \mbox{Ran}(Z)\}$ is continuous.
The following Theorems 3.1-3.3 give the axiomatic characterization of normalized CDRMs under random environment,
which are the main results of the present paper.

\vspace{0.2cm}

\noindent{\bf Theorem 3.1}\ \  Suppose that Assumption A holds.
If a normalized conditional risk measure under random environment $\rho: \sX \times \sX^\perp  \rightarrow \mathbb{R}$  satisfies
Axioms B2-B4, then for any random environment $Z \in \sX^\perp $, there is a $P_Z-$null set $N \in \sB(\mathbb{R})$,
so that there exist a family $\{g_z; z \in N^c\cap\mbox{Ran}(Z)\}$ of
left-continuous distortion functions depending on the states of $Z$ and a monotone, normalized set function $\gamma_Z$ on $\sB(\mbox{Ran}(Z))$
depending on $Z$ such that for any $X \in \sX_+$,
$$
\rho(X; Z)
   = \int_0^{\infty} \gamma_Z \left( \left \{ z \in N^c\cap\mbox{Ran}(Z) :
                                      \int_0^{\infty} g_z \circ K_Z(z,\{ X > \alpha \})d\alpha >\beta \right \} \right) d\beta,
$$
where the $P_Z-$null set $N \in \sB(\mathbb{R})$ and the distortion functions $g_z$ are given as in Proposition 3.3.

\vspace{0.2cm}

Besides the conditions as in Theorem 3.1, if a normalized conditional risk measure under random environment is assumed to
further satisfy Axiom B1, then the monotone, normalized set function
$\gamma_Z$ on $\sB(\mbox{Ran}(Z))$ as in Theorem 3.1 can be further expressed in terms of a distorted probability,
which leads to the following Theorem 3.2.

\vspace{0.2cm}

\noindent{\bf Theorem 3.2}\ \ Suppose that Assumption A holds.
If a normalized conditional risk measure under random environment $\rho: \sX \times \sX^\perp  \rightarrow \mathbb{R}$ satisfies
Axioms  B1-B4, then for any random environment $Z \in \sX^\perp $, there is a $P_Z-$null set $N \in \sB(\mathbb{R})$,
so that there exist a family $\{g_z; z \in N^c\cap\mbox{Ran}(Z)\}$ of left-continuous
distortion functions depending on the states of $Z$ and a function $h_Z : [0,1] \rightarrow [0,1]$ depending on $Z$ with $h_Z(0)$ $=$ $0$
and $h_Z(1) = 1$ such that for any $X \in \sX_+$,
$$
\rho(X;Z)
= \int_0^{\infty} h_Z \circ P_Z^* \left( \left \{ z \in N^c\cap\mbox{Ran}(Z) :
                                           \int_0^{\infty} g_z \circ K_Z(z,\{ X > \alpha \}) d\alpha > \beta \right \} \right) d\beta,
$$
where the $P_Z-$null set $N \in \sB(\mathbb{R})$ and the distortion functions $g_z$ are as in Theorem 3.1.

\vspace{0.2cm}

Next theorem extends Theorem 3.2 to general random losses, and thus gives axiomatic characterization for
a normalized CDRM under random environment.

\vspace{0.2cm}

\noindent{\bf Theorem 3.3}\ \ Suppose that Assumption A holds.
If a normalized conditional risk measure under random environment $\rho: \sX \times \sX^\perp  \rightarrow \mathbb{R}$  satisfies
Axioms  B1-B4, then for any random environment $Z \in \sX^\perp $, there is a $P_Z-$null set $N \in \sB(\mathbb{R})$,
so that there exist a family $\{g_z; z\in N^c\cap\mbox{Ran}(Z)\}$ of
left-continuous distortion functions depending on the states of $Z$ and a function $h_Z : [0,1] \rightarrow [0,1]$ depending on $Z$
with $h_Z(0)$ $ = $ $0$ and $h_Z(1) = 1$ such that for any $X \in \sX$,
\begin{align}\label{3002}
\rho (X; Z)
  & = \int_{-\infty}^0 \left[ h_Z \circ P_Z^* \left( \left\{ z \in N^c\cap\mbox{Ran}(Z) : \tau_Z(X; z) > \beta \right\} \right) - 1 \right] d\beta \nonumber\\
  & \quad + \int_0^{\infty} h_Z \circ P_Z^* \left( \left\{ z \in N^c\cap\mbox{Ran}(Z) : \tau_Z(X; z) > \beta \right\} \right) d\beta,
\end{align}
where the $P_Z-$null set $N \in \sB(\mathbb{R})$ and the function $h_Z$ are as in Theorem 3.2,
and for $z \in N^c\cap\mbox{Ran}(Z),$ $\tau_Z(\cdot; z) $ as in Assumption A is given by
\begin{align*}
\tau_Z(X; z) = \int_{-\infty}^0 \left[ g_z \circ K_Z(z,\{ X > \alpha \})- 1 \right]  d\alpha
                + \int_0^{\infty} g_z \circ K_Z(z,\{ X > \alpha\}) d\alpha,
\end{align*}
where the distortion functions $g_z$ are given as in Theorem 3.2.

\subsection{Coherence and dual presentation}

In this subsection, we discuss the coherence and dual representation for the normalized CDRM under random environment $\rho$
defined by $(\ref{add3002})$ with respect to the first input argument. More precisely, provided with a random environment $Z \in \sX^\perp $,
considering $\rho_{_Z}(\cdot) := \rho (\cdot; Z) : \sX \rightarrow \mathbb{R}$ as a \textit{univariate} functional of random loss,
we discuss its coherence and dual representation.

\vspace{0.2cm}

In general, a risk measure is defined as any functional $\rho : \sX \rightarrow \mathbb{R}$.  Artzner et al. (1999) initiated coherent risk measures.
A risk measure $\rho : \sX \rightarrow \mathbb{R}$ is called coherent, if it satisfies the following four properties (axioms):
\begin{enumerate}
  \item[(i)] Monotonicity : $X\leq Y$ implies $\rho(X)\leq \rho(Y)$ for any $X, Y\in \sX$.

  \item[(ii)] Translation invariance : $\rho(X+a)=\rho(X)+a$ for any $X\in \sX$ and $a\in \mathbb{R}$.

  \item[(iii)] Positive homogeneity : $\rho(cX)=c\rho(X)$ for each $c > 0$ and $X\in \sX$.

  \item[(iv)] Subadditivity : $\rho(X+Y)\leq \rho(X)+\rho(Y)$ for any $X, Y \in \sX$.

\end{enumerate}

\vspace{0.2cm}

A risk measure $\rho : \sX \rightarrow \mathbb{R}$ is called normalized, if $\rho (1) = 1 $,
and is called comonotonically additive, if $\rho(X+Y) = \rho(X) + \rho(Y)$ whenever $X$ and $Y$ are comonotonic on $\Omega$.
Notice that a normalized risk measure $\rho$ is translation invariant (also known as cash invariant),
if it is positively homogeneous and comonotonically additive. In fact, positive homogeneity implies that $\rho(0)=0$.
Hence by normalization and comonotonic additivity, $0 = \rho (1-1) = \rho(1) + \rho(-1)$, which yields that $\rho(-1) = -1$.
Therefore, we have that $\rho(a) = a$ for any $a\in \mathbb{R}$, which, together with the comonotonic additivity, implies
the translation invariance.

\vspace{0.2cm}

Now, we state the first main result of this subsection, which concerns the coherence of the functional $\rho_{_Z}.$

\vspace{0.2cm}

\noindent{\bf Theorem 3.4}\ \ Let $ \{g_z; z \in \mathbb{R} \} $ be a family of concave distortion functions, which is regular on $\mathbb{R}.$
Let a random environment $Z \in \sX^\perp $ be fixed, and $h_Z$ be a concave distortion function associated with $Z$.
Then the risk measure $\rho_{_Z} : \sX \rightarrow \mathbb{R}$ defined by
\begin{align}\label{3001}
 \rho_{_Z}(X)
      & :=  \int_{-\infty}^0 \left[ h_Z \circ P_Z \left( \left\{ z: \int_{-\infty}^0 \left( g_z \circ K_Z(z, \{X > \alpha \})- 1\right)d\alpha
                             \right. \right. \right. \nonumber \\
      & \quad + \left. \left. \left. \int_0^{\infty} g_z \circ K_Z(z, \{X > \alpha \})d\alpha >\beta \right\} \right) - 1 \right] d\beta \nonumber \\
      & \quad + \int_0^{\infty} h_Z \circ P_Z \left( \left\{ z: \int_{-\infty}^0 \left( g_z \circ K_Z(z, \{ X > \alpha \})- 1 \right) d\alpha
                                              \right. \right. \nonumber\\
      & \quad + \left. \left. \int_0^{\infty} g_z \circ K_Z(z,\{ X > \alpha \}) d\alpha > \beta \right\} \right) d\beta, \quad X \in \sX,
\end{align}
 is coherent.

\vspace{0.2cm}

Next, we turn to discuss the dual representation for $\rho_{_Z}$. It turns out that $\rho_{_Z}$
can be expressed by means of a repeated Choquet integral on some product space with respect to two finitely additive
probability measures. For this purpose, we need a little more preparations.

\vspace{0.2cm}

Given a monotone set function $\mu_1 : \sB(\mathbb{R}) \rightarrow \mathbb{R}_+$ and a set $ \mu_2 := \{\mu_2(z,\cdot); z\in \mathbb{R} \}$
of monotone set functions on $\sF$ satisfying that $ \mu_2(\cdot, A )$ is a Borel function on $ \mathbb{R}$
for every $A \in \sF$,
we define a set function $ \mu $ on  $ \sB(\mathbb{R}) \times \sF$, the product $\sigma$-algebra of $ \sB(\mathbb{R})$ and $ \sF$, through
\begin{align*}
  \mu(A):= \int \left(\int 1_{A}(z,\omega){\mu_2}(z,d\omega) \right) \mu_1(dz), \quad A \in  \sB (\mathbb{R})\times \sF,
\end{align*}
where the repeated integral is understood as Choquet integral.
Apparently, $ \mu $ is monotone; for instance, see Proposition 12.1(i) of
Denneberg (1994). Such a defined $\mu$ is called the generalized product of the system $ \{ \mu_1, \mu_2(z, \cdot); z \in \mathbb{R} \}$,
and is denoted by $\mu := \mu_1 \otimes \mu_2.$
For a bounded random variable $Y $ on $ (\mathbb{R}\times \Omega, \sB(\mathbb{R})\times \sF)$, the Choquet integral of $Y$
with respect to $\mu$ is denoted by $ \int Y d\mu.$  If $\mu_1$ is finitely additive (i.e. $ \mu_1 (A \cup B ) = \mu_1 (A) + \mu_1 (B) $ for disjoint
sets $A$ and $B$), then an application of Proposition 12.1(iv) of Denneberg (1994) implies that
\begin{align*}
  \int Y d\mu = \int \left(\int Y(z,\omega){\mu_2}(z,d\omega) \right) \mu_1(dz).
\end{align*}
In the sequel, in order to avoid tedious measurability considerations, it is sometimes convenience and helpful
for us to extend a monotone set function on $ \sB(\mathbb{R}) $ onto the power set $ 2^\mathbb{R},$ the family of all subsets of
$ \mathbb{R}$. Let $\nu : \sB(\mathbb{R}) \rightarrow \mathbb{R}_+ $ be a monotone set function on $  \sB(\mathbb{R}) $. Define
\begin{align*}
  \widehat{\nu} (A) := \inf \{ \nu(B) : A \subseteq B \in \sB(\mathbb{R}) \}, \ \ A \in 2^\mathbb{R}.
\end{align*}
The set function $ \widehat{\nu} : 2^\mathbb{R} \rightarrow \mathbb{R}_+ $ is called outer set function of $\nu$. Apparently,
$ \nu = \widehat{\nu} $ on $ \sB(\mathbb{R}).$ Furthermore, it is known that $ \widehat{\nu} $ is monotone on $2^\mathbb{R}$, see Proposition 2.4(i)
of Denneberg (1994). Under this point of view, in the course of defining the generalized product $\mu = \mu_1 \otimes \mu_2 $,
we could replace $ \mu_1 $ with its outer set function  $ \widehat{\mu}_1, $ and drop the assumption on $\mu_2$ that $ \mu_2(\cdot, A )$ is a Borel
function on $ \mathbb{R}$ for every $A \in \sF$. Then we would define a generalized product $\widehat{ \mu} := \widehat{\mu}_1 \otimes \mu_2 $ on
$ 2^\mathbb{R} \times \sF, $ the product $\sigma$-algebra of $ 2^\mathbb{R}$ and $ \sF$. In this situation, we could
similarly define the Choquet integral $ \int Y d \widehat{\mu}$ of a bounded random variable $Y$ on $ (\mathbb{R} \times \Omega, 2^\mathbb{R} \times \sF)$
with respect to $ \widehat{ \mu }$, and all relevant conclusions still remain true. Particularly,
$ \int Y d \widehat{\mu} = \int Y d \mu $ for all bounded random variables $Y$ on $ (\mathbb{R} \times \Omega, \sB(\mathbb{R}) \times \sF).$
For more details, we refer to Chapters 2 and 12 of Denneberg (1994).

\vspace{0.2cm}

We introduce more notations. $ \sM_{1,f}(\Omega, \sF)$ denotes the set of all finitely additive normalized set functions
$Q : \sF  \rightarrow [0,1]$, and $E_Q(X)$ denotes the integral of $X \in \sX$ with respect to $Q$,
as constructed in Theorem A.54 of F{\"{o}}llmer and Schied (2016).  An application of Lemma 4.97 of F{\"{o}}llmer and Schied (2016) yields
that the integral $ E_Q(X) $ is equal to the Choquet integral $\int X dQ.$
Similarly, $ \sM_{1,f}(\mathbb{R}, \sB(\mathbb{R})) $ denotes the set of all finitely additive normalized set
functions $Q : \sB(\mathbb{R}) \rightarrow [0,1]$, and $E_Q(W)$ denotes the integral of a Borel function $W$ on $ \mathbb{R}$ with respect to $Q$.
Again, applying Lemma 4.97 of F{\"{o}}llmer and Schied (2016) to the measurable space $ ( \mathbb{R}, \sB(\mathbb{R}))$ implies that the
integral $E_Q(W)$ is equal to the Choquet integral $ \int W dQ$ for bounded Borel function $W$ on $\mathbb{R}.$

\vspace{0.2cm}

Now, we are ready to state the second main result of this subsection, which gives the dual representation for $\rho_{_Z}$ defined by $(\ref{3001})$,

\vspace{0.2cm}

\noindent{\bf Theorem 3.5}\ \ Let $ \{g_z; z \in \mathbb{R} \}$ be a family of concave distortion functions, which is regular on $\mathbb{R}.$
Let a random environment $Z \in \sX^\perp $ be fixed, and $h_Z$ be a concave distortion function associated with $Z$.
Denote $ \sQ_1 := \{Q_1 \in \sM_{1,f}(\mathbb{R}, \sB(\mathbb{R})) : Q_1(B) \leq h_Z \circ P_Z(B)\ \mbox{for\ all}\ B \in \sB(\mathbb{R})\},$
$\sQ_2 := \{ Q_2 := \{ Q_2(z, \cdot) \in \sM_{1,f}(\Omega, \sF); z \in \mathbb{R} \} :  \mbox{for\ every}\ z \in \mathbb{R}, \
 Q_2(z, A) \leq g_z \circ K_Z(z, A)\ \mbox{for\ all}\ A \in \sF \}.$
Let $\sC := \{ (Q_1, Q_2) : Q_1 \in \sQ_1, Q_2 \in \sQ_2 \}.$
Then for any random loss $ X \in \sX$,
\begin{align}\label{add3013}
  \rho_{_Z}(X)
    = \underset{(Q_1, Q_2) \in \sC}{\sup} \int \left( \int X(\omega) Q_2(z, d\omega) \right) \widehat{Q}_1(dz),
\end{align}
where $\widehat{Q}_1$ is the outer set function of $Q_1$, and the supremum taken over $ \sC $ can be attained.

\vspace{0.2cm}

\noindent{\bf Remark 3.4}\ \  If the random environment $Z$ is degenerate, say $P( Z = z_0) = 1 $ with some $z_0 \in \mathbb{R}$,
then for each $z \in \mathbb{R}$, $K_Z(z,\cdot) = P(\cdot)$, and $ \sQ_1 $ is the singleton $\{ \delta_{z_0} \}$ consisting of
the Dirac measure concentrating at $z_0$. Hence, ($\ref{add3013}$) reduces to
\begin{align*}
  \rho_{z_0}(X)
    = \underset{Q(z_0, \cdot) \in \sQ}{\sup}  \int X(\omega) Q(z_0, d\omega)
    = \underset{Q(z_0, \cdot) \in \sQ}{\sup}  E_{ Q(z_0, \cdot)}[X],
\end{align*}
where $\sQ := \{ Q(z_0, \cdot) \in \sM_{1,f}(\Omega, \sF) : Q(z_0, A) \leq g_{z_0} \circ P(A)\ \mbox{for\ all}\ A \in \sF \}.$
This just coincides with the classical representation result; for instance, see Theorem 4.94(c) and (d) of F{\"{o}}llmer and Schied (2016).

\subsection{Distortion risk measures in the presence of background risk}

Tsanakas (2008) established a distortion risk measure in the presence of background risk.
In relation to background risk, see also Gollier and Pratt (1996) and Heaton and Lucas (2000).
In this subsection, by making use of the CDRM under random environment defined by $(\ref{add3002})$, we will introduce two new risk measures
in the presence of background risk. More precisely, by letting a random environment be the sum of a given random loss and the background risk,
we will establish two new distortion risk measures in the presence of background risk. It turns out that they bound the distortion risk measure
introduced by Tsanakas (2008) from below and above, respectively.

\vspace{0.2cm}

Throughout this subsection, let $g : [0,1] \rightarrow [0,1]$ be an increasing,
concave and differentiable function with $g(0)=0$, $g(1)=1$  and bounded first derivative $g'$. Let $Y$ be a random variable on $(\Omega, \sF, P)$
representing a background risk. Let $X$ be a random variable on $(\Omega, \sF, P)$ representing a random loss.
For the convenience of our discussion and without loss of generality, we assume that $\sX^\perp  = \sX$ and
$X, Y \in \sX$. Define $Z := X + Y$ serving as a random environment, $\overline{F}_{Z}(z) := P(Z >z)$, $z \in \mathbb{R}$, and a distribution
function $ s(z):= 1-g(\overline{F}_Z(z))$, $z \in \mathbb{R}.$  Let $L_s$ be the Lebesgue-Stieltjes measure induced by $s(z)$ on $\sB(\mathbb{R})$.
Let $\phi : \mathbb{R} \rightarrow \mathbb{R}$ be a Borel function such that $E[X|Z] = \phi(Z)$ and $ E[X|Z = z] = \phi(z)$ for any $z \in \mbox{Ran}(Z)$.

\vspace{0.2cm}

Now, making use of $(\ref{add3002})$, we turn to construct a conditional distortion risk measure of $X$ under random environment $Z$. For simplicity,
for any $z \in \mbox{Ran}(Z)$, let the distortion function $g_z$ in $(\ref{add3002})$ be the identity function, that is, $g_z(x) = x$,
and let the distortion function $h := h_Z$ in $(\ref{add3002})$ be chosen later.
Thanks to $(\ref{add3002})$, we define a risk measure
$\rho_{_h}(X; Z)$ of $X$ under random environment $Z$ by
\begin{align}\label{3003}
\rho_{_h}(X; Z)
  & := \int_{-\infty}^0 \left[ h_Z \circ P_Z \left( \left\{ z: \int_{-\infty}^0 \left( K_Z(z,\{ X > \alpha \})- 1\right) d\alpha
                        \right. \right. \right. \nonumber \\
  & \quad + \left. \left. \left. \int_0^{\infty}  K_Z(z,\{ X > \alpha \}) d\alpha > \beta \right\} \right) - 1 \right] d\beta \nonumber \\
  & \quad + \int_0^{\infty} h_Z \circ P_Z \left( \left\{ z: \int_{-\infty}^0 \left( K_Z(z,\{ X > \alpha \})- 1\right) d\alpha \right. \right. \nonumber \\
  & \quad + \left. \left. \int_0^{\infty} K_Z(z,\{ X > \alpha \}) d\alpha > \beta \right\} \right) d\beta.
\end{align}
We call $\rho_{_h}(X; Z)$ the distortion risk measure of $X$ with respect to background risk $Y.$
Note that by ($\ref{0415add1}$),
\begin{align*}
 \int_{-\infty}^0 & (K_Z(z,\{ X > \alpha \})-1)d\alpha+ \int_0^{\infty} K_Z(z,\{ X > \alpha \})d\alpha \\
                  & = \int_\Omega X(\omega) K_Z(z, d\omega) = E[X|Z = z] = \phi (z) \quad \mbox{for}\ P_Z-a.e.\ z \in \mathbb{R},
\end{align*}
which, along with $(\ref{3003}),$ yields an alternative expression for $\rho_{_h}(X; Z):$
\begin{align}\label{add3006}
\rho_{_h}(X; Z) = \int_{-\infty}^0 \left[ h_Z \circ P_Z (\phi > \beta) - 1 \right ] d\beta
               + \int_0^{\infty} h_Z \circ P_Z (\phi > \beta )  d\beta.
\end{align}

\vspace{0.2cm}

Next, we will specify the distortion function $h_Z$. For this purpose, we define two functions
$v, u : \mathbb{R} \rightarrow [0, 1]$ by
\begin{align*}
v(x) := L_s(\{z: \phi(z) > x \}), \ x\in \mathbb{R}, \\
u(x) := P_Z(\{ z: \phi(z) > x \}), \ x\in \mathbb{R}.
\end{align*}
Then $v$ and $u$ are non-increasing and right-continuous functions, and take values in $[0,1].$
We denote by $u^{-1}$ and $u^{-1+}$  the left-continuous and right-continuous inverse functions of $u$,
respectively, that is, for any $p \in (0,1)$,
\begin{align*}
  u^{-1}(p):= \inf\{ x \in \mathbb{R} : \ u(x) \leq p \},\\
  u^{-1+}(p):= \sup\{ x \in \mathbb{R} : \ u(x) \geq p\},
\end{align*}
with $u^{-1}(0) :=  u^{-1+}(0) := +\infty$ and $u^{-1}(1) :=  u^{-1+}(1) := -\infty$ by convention.
Note that  $u^{-1}(p)$ and $ u^{-1+}(p)$ are finite for all $ p\in (0,1).$  Also clearly,
\begin{align*}
 u^{-1}(p)  \leq u^{-1+}(p) \quad \mbox{for\ any\ } p\in (0,1).
\end{align*}
Note that let $x \in \mathbb{R}$ be such that $0 < u(x) <1$, then $u^{-1}(u(x))$ and $ u^{-1+}(u(x))$ are finite, and
\begin{align}\label{add3008}
u^{-1}(u(x)) \leq x \leq u^{-1+}(u(x)).
\end{align}
For more details about different inverse functions of $u,$ we refer to Dhaene et al. (2002) and F\"{o}llmer and Schied (2016, Appendix A.3).

\vspace{0.2cm}

Define two functions $h_L$ and $h_R$ on $[0,1]$, respectively, by
\begin{align*}
  h_L(p):= v[u^{-1}(p)], \quad p \in [0,1], \\
  h_R(p):= v[u^{-1+}(p)], \quad p \in [0,1].
\end{align*}
Clearly, $h_L$ and $h_R$ are two distortion functions.
We denote by $\rho_{_L}$ and $\rho_{_R}$ the distortion risk measure $ \rho_{_h} $ as in $(\ref{add3006})$, when
the distortion function $h_Z$ is chosen as $h_L$ and $h_R$, respectively.   Note that $v$ is non-increasing, from $(\ref{add3008})$
and the definitions of $h_L$ and $h_R$,  it follows that for any $\beta \in \mathbb{R},$
\begin{align}\label{0418add1}
  h_L(u(\beta)) = v[u^{-1}(u(\beta))] \geq v(\beta)  \geq v[u^{-1+}(u(\beta))] = h_R(u(\beta)),
\end{align}
which, together with $(\ref{add3006})$, results in the following proposition:

\vspace{0.2cm}

\noindent{\bf Proposition 3.5}\ \ It holds that
\begin{align*}
\rho_{_R}(X; Z)  \leq \rho_{_L}(X; Z).
\end{align*}

\vspace{0.2cm}

Let us end this subsection with a comparison with the distortion risk measure introduced by Tsanakas (2008).
In Tsanakas (2008),  the distortion risk measure $\Gamma (X;Y)$ of risk $X$ with respect to
background risk $Y$ is defined by
\begin{align}\label{add3003}
  \Gamma(X; Y) := E[Xg'(\overline{F}_{Z}(Z))],
\end{align}
where $Z := X+Y.$  By the total expectation law,
\begin{align}\label{add3004}
  E[Xg'(\overline{F}_{Z}(Z))]
     = E[E(X|Z)g'(\overline{F}_Z(Z))]
     = \int_{-\infty}^{\infty} \phi(z)g'(\overline{F}_Z(z))F_{Z}(dz).
\end{align}
By Fubini's theorem, an elementary calculation yields that
\begin{align*}
 \int_{-\infty}^{\infty} \phi(z)g'(\overline{F}_Z(z))F_{Z}(dz)
     = \int_{-\infty}^0 \left[ L_s \left( \phi > \beta \right)-1 \right] d\beta
             + \int_0^{\infty} L_s \left( \phi > \beta \right) d\beta,
\end{align*}
which, together with $(\ref{add3003})$ and $(\ref{add3004})$, implies
\begin{align}\label{add3005}
  \Gamma(X; Y)
      = \int_{-\infty}^0 \left[ L_s \left( \phi > \beta \right)-1 \right] d\beta
             + \int_0^{\infty} L_s \left( \phi > \beta \right) d\beta.
\end{align}
Consequently, from $(\ref{add3006})$, $(\ref{0418add1})$ and  $(\ref{add3005})$, it follows that
\begin{align*}
\rho_{_R}(X; Z) \leq \Gamma(X; Y) \leq \rho_{_L}(X; Z).
\end{align*}

\section{Examples}

In this section, by examples, we illustrate the introduced CDRMs under random environment.
In the first subsection, we will show that under certain mild condition, the class of CDRMs under random environment
includes the WVaR and RVaR as special cases, where the random environment can describe an institution's risk preference.
In the second subsection, we will address the connection between the CDRMs under random environment and
$\sQ$-mixtures of ES, where the random environment can describe different probability measures (scenarios) representing model uncertainty.

\subsection{WVaR and RVaR}

Throughout this subsection, for simplicity, we assume that $\sX^\perp  = \sX$. Let us start with recalling the definitions of WVaR
introduced by Acerbi (2002) and Cherny (2006), and RVaR studied by Cont et al. (2010) and Embrechts et al. (2018).

\vspace{0.2cm}

Given a probability measure $\nu$ on $(0,1)$, the WVaR of a random loss $X \in \sX$ with respect to $\nu$
is defined by
\begin{align*}
  \mbox{WVaR}_{\nu}(X) := \int_{(0,1)} \mbox{ES}_{\theta}(X)\nu(d\theta).
\end{align*}
Given two confidence levels $0 < \alpha_1 < \alpha_2 < 1$, the RVaR of a random loss $X \in \sX$ cross the confidence level range
$[\alpha_1, \alpha_2]$ is defined by
\begin{align*}
  \mbox{RVaR}_{\alpha_1,\alpha_2}(X) := \frac{1}{\alpha_2-\alpha_1}\int_{\alpha_1}^{\alpha_2} \mbox{VaR}_{\theta}(X) d\theta.
\end{align*}

\vspace{0.2cm}

Let $\mu$ be a probability measure on $(0, 1)$. Denote by $F_\mu$ the distribution function induced by $\mu$, that is,
\begin{align*}
  F_\mu (x)
    :=\begin{cases}
       \mu((0,1) \cap (0, x]), \quad\quad\quad & \mbox{if}\quad  x > 0, \\
       0, \quad\quad\quad & \mbox{if}\quad x \leq 0.
\end{cases}
\end{align*}
Denote by $F_\mu^{-1+}$ the right-continuous inverse function of $F_\mu,$ that is $F_\mu^{-1+}(p) := \inf \{x \in \mathbb{R} : F_\mu(x) > p\},$
$p \in (0,1);$ $F_\mu^{-1+}(0) := 0,$ $F_\mu^{-1+}(1) := 1.$

\vspace{0.2cm}

Given a random loss $X \in \sX,$ by the assumption that $\sX^\perp  = \sX$, let $U$ be a U[0, 1] random variable on $(\Omega, \sF, P)$
independent of $X$. Define a random environment $Z := F_\mu^{-1+}(U)$, then $Z$ is independent of $X$ and $P_Z^* = \mu.$
Denote by $N_0$ the $P_Z-$null set as in ($\ref{0429add1}$). Hence, by ($\ref{0415add1}$) we know that there exists a $P_Z-$null set
$N \in \sB(\mathbb{R})$ with $N_0 \subseteq N$ such that for every $z\in N^c$ and any $t \in \mathbb{R}$,
\begin{align}\label{add4001}
 K_Z(z, \{X>t\}) = E[1_{\{X>t\}} | Z=z] = P(X>t).
\end{align}
For $Z$ as above, define a continuous distortion function $h_Z$ associated with $Z$ as the identity function, that is,
\begin{align}\label{add4004}
    h_Z(x) := x,\ \  x \in [0,1].
\end{align}
Next, we proceed by two cases:

\vspace{0.2cm}

{\bf Case 1:} For each $z \in (0,1)$, we define a
concave distortion function $g_z : [0,1] \rightarrow [0,1]$ by
\begin{align}\label{add4002}
  g_z(x)
    :=\begin{cases}
       \frac{x}{1-z}, \quad\quad\quad & \mbox{if}\quad 0\leq x\leq 1-z, \\
       1, \quad\quad\quad & \mbox{if}\quad 1-z< x \leq 1,
\end{cases}
\end{align}
and, for $z \notin (0,1)$, we define $g_z : [0,1] \rightarrow [0,1]$ by $ g_z(x) := x. $
Clearly, the family $\left\{ g_z; z \in \mathbb{R} \right\}$ is regular on $\mathbb{R}$, and thus it is also regular on $(0,1)$.
It is well-known that to the distortion function $g_z$ as in ($\ref{add4002}$), the corresponding distortion risk measure is exactly the ES,
for example, see Belles-Sampera et al. (2014), F\"{o}llmer and Schied (2016, Example 4.71), or Wang and Ziegel (2021).

\vspace{0.2cm}

For each $z \in [0,1]$, let the risk measure $\rho_{_Z}(\cdot;z) : \sX \rightarrow \mathbb{R}$ be as in ($\ref{add3001}$),
and the risk measure $\rho : \sX \times \sX^\perp   \rightarrow \mathbb{R}$ as in ($\ref{add3002}$).
Noting ($\ref{add4001}$), after plugging distortion functions $g_z$ as above into ($\ref{add3001}$), we have that for every $z\in N^c \cap \mbox{Ran}(Z)$,
\begin{align}\label{add4005}
  \rho_{_Z}( X; z)
         =  \int_{-\infty}^0 (g_z \circ P( X > \alpha )-1)d\alpha
             + \int_0^{\infty} g_z \circ P( X > \alpha )d\alpha
         =  \mbox{ES}_{z}(X).
\end{align}

\vspace{0.2cm}

Recall that the Choquet integral of $\rho_{_Z}(X; \cdot)$ with respect to the probability measure $P_Z^*$ is equal to the integral
of $\rho_{_Z}(X; \cdot)$ with respect to $P_Z^*$, see Lemma 4.97 of F\"{o}llmer and Schied (2016).
Consequently, substituting ($\ref{add4004}$) into ($\ref{add3002}$), by ($\ref{add3012}$) and ($\ref{add4005}$) we have that
\begin{align*}
  \rho (X; Z)
         = \int \rho_{_Z}(X; z) P_Z^*(dz)
         = \int_{(0,1)} \mbox{ES}_z(X) \mu(dz),
\end{align*}
which is just the $\mbox{WVaR}_{\mu}(X)$.

\vspace{0.2cm}

{\bf Case 2:} Given two confidence levels $0 < \alpha_1 < \alpha_2 < 1$, suppose that the probability measure $\mu$ has a probability density function
\begin{align*}
f_{\mu}(x) := \frac{1}{\alpha_2- \alpha_1} 1_{[\alpha_1,\ \alpha_2]}(x),\quad \quad x \in (0,1).
\end{align*}
Define a family $\left\{ g_z; z \in \mathbb{R} \right\}$ of left-continuous distortion functions by
\begin{align*}
    g_z(x)
  :=\begin{cases}
  0, \quad\quad\quad & \mbox{if}\quad 0\leq x\leq 1-z, \\
  1, \quad\quad\quad & \mbox{if}\quad 1-z< x \leq 1,
  \end{cases}
\end{align*}
when $z \in [\alpha_1, \alpha_2]$, while $g_z(x) := x, x \in [0,1],$  when $z \notin [\alpha_1, \alpha_2]$.
Apparently, $\left\{ g_z; z \in \mathbb{R} \right\}$ is regular on $\mathbb{R}$.
It is well-known that to the distortion function $g_z, z \in [\alpha_1, \alpha_2],$  the corresponding distortion risk measure
is exactly the VaR, for example, see Belles-Sampera et al. (2014) or Wang and Ziegel (2021).

\vspace{0.2cm}

For any $z \in [\alpha_1,\alpha_2]$, again let the risk measure $\rho_{_Z}(\cdot;z) : \sX \rightarrow \mathbb{R}$
be as in ($\ref{add3001}$), and the risk measure $\rho : \sX \times \sX^\perp   \rightarrow \mathbb{R}$  as in ($\ref{add3002}$).
Noting ($\ref{add4001}$), after substituting distortion functions $g_z$ as above into ($\ref{add3001}$)
we know that for every $z\in N^c \cap \mbox{Ran}(Z),$
\begin{align*}
\rho_{_Z}( X; z)
         =  \int_{-\infty}^0 (g_z \circ P( X > \alpha )-1)d\alpha
             + \int_0^{\infty} g_z \circ P( X > \alpha )d\alpha
         =  \mbox{VaR}_{z}(X).
\end{align*}
Consequently, by ($\ref{add3002}$), ($\ref{add3012}$) and Lemma 4.97 of F\"{o}llmer and Schied (2016), we know that
\begin{align*}
    \rho( X; Z)
           = \int \rho_{_Z}(X; z) P_Z^*(dz)
           = \int_{[\alpha_1, \alpha_2]} \mbox{VaR}_z(X) \mu(dz)
           = \frac{1}{\alpha_2-\alpha_1} \int_{\alpha_1}^{\alpha_2} \mbox{VaR}_{\theta}(X) d\theta,
\end{align*}
which is just the $ \mbox{RVaR}_{\alpha_1, \alpha_2}(X)$.

\subsection{$\sQ$-mixture of ES}

For mutually singular probability measures (scenarios) $Q_1, \cdots, Q_n$ on the measurable space $(\Omega, \sF),$ Wang and Ziegel (2021) showed that
if $(\Omega, \sF, Q_i)$ is atomless for each $i=1, \cdots, n,$ then a risk measure
is a $\sQ$-based coherent risk measure if and only if it is a supremum of $\sQ$-mixtures of ES; for more details, see Wang and Ziegel (2021).
In this subsection, we address the connection of CDRM under random environment with $\sQ$-mixture of ES. We concentrate on the way of how the
CDRM under random environment is applied to risk measures with model uncertainty. Let us briefly recall the definition of a $\sQ$-mixture of ES.

\vspace{0.2cm}

Let $Q_1, \cdots, Q_n$ be $n$ probability measures (scenarios) on the  measurable space $(\Omega, \sF),$ which are mutually singular, that is, there are
disjoint sets $\Omega_1, \cdots, \Omega_n \in \sF$ with $Q_i(\Omega_i) = 1$ for $i=1, \cdots, n,$ and $ \cup_{i=1}^{n} \Omega_i = \Omega.$
A $Q$-mixture of ES is a risk measure $\hat{\rho}$ defined by
\begin{align*}
\hat{\rho}(X) := \sum^n_{i=1} w_i \int^1_0 \mbox{ES}^{Q_i}_p(X)\mbox{d}h_i(p), \quad X\in \sX,
\end{align*}
for some $(w_1, \cdots, w_n) \in [0,1]^n$ with $\sum^n_{i=1} w_i =1$ and distribution functions $h_1, \cdots, h_n$ on $[0,1];$
see Wang and Ziegel (2021). Clearly, without loss of generality, we can also assume that $(w_1, \cdots, w_n) \in (0,1)^n.$

\vspace{0.2cm}

Next, we proceed to show how to deduce a $Q$-mixture of ES by means of the CDRM under random environment
defined by (\ref{add3002}).
Let $Q_1, \cdots, Q_n$ be $n$ mutually singular probability measures (scenarios) on $(\Omega, \sF)$ such that $(\Omega, \sF, Q_i)$ is atomless
for each $i=1, \cdots, n.$
Let $\Omega_1, \cdots, \Omega_n \in \sF$ be disjoint sets with $Q_i(\Omega_i) = 1$ for $i=1, \cdots, n,$ and $ \cup_{i=1}^{n} \Omega_i = \Omega.$
Let a $(w_1, \cdots, w_n) \in (0,1)^n$ with $\sum^n_{i=1} w_i =1$ and distribution functions $h_1, \cdots, h_n$ on $[0,1]$
be given. Define a reference probability measure $\tilde{P}$ on $(\Omega, \sF)$ by
\begin{align*}
\tilde{P}(A) := \sum^n_{i=1} w_iQ_i(A), \quad A\in \sF.
\end{align*}
Relevant symbol of expectation with respect to $\tilde{P}$ is denoted by $\tilde{E}$.
Define a discrete random variable $Z : \Omega \rightarrow \mathbb{R}$ by $Z(\omega) := i$ if $\omega \in \Omega_i.$
Clearly, the probability mass function of $Z$ under $\tilde{P}$ is given by $\tilde{P}(Z=i) = \omega_i$ for each $i=1, \cdots, n.$
Moreover, $Z\in \sX^\perp$. It is not difficult to verify that under $\tilde{P},$ the regular conditional probability
$\{ K_Z(i,\cdot) : i\in \{1, \cdots, n\} \}$ with respect to $Z$ is given by
\begin{align*}
 K_Z(i, A) = \tilde{E}[1_A| Z=i] = Q_i(A), \quad A\in \sF,
\end{align*}
for each $i=1, \cdots, n.$  For each $h_i, i=1, \cdots, n,$ we associate it with a distortion function $g_i$ defined by
\begin{align*}
g_i(x) := \int^1_0 \int_{(t,1]}s^{-1}\mbox{d}h_i(s)\mbox{d}t, \quad x\in [0,1].
\end{align*}
Then, under $Q_i, i=1, \cdots, n,$ the distortion risk measure of a random loss $X\in \sX$ with distortion function $g_i$ is a WVaR.
More precisely,
\begin{align*}
\int X \mbox{d}g_i \circ Q_i = \int^1_0 \mbox{ES}^{Q_i}_p(X)\mbox{d}h_i(p), \quad X\in \sX;
\end{align*}
for instance, see F\"{o}llmer and Schied (2016, Theorem 4.70). We also associate $Z$ with an identity distortion function $h_Z$, that is,
$h_Z(x) := x, x\in [0,1]$.
Therefore, by Definition 3.2, we know that for each $i=1, \cdots, n,$ the CDRM under fixed state $i$ is given by
\begin{align*}
 \rho_{_Z}(X; i) =  \int^1_0 \mbox{ES}^{Q_i}_p(X)\mbox{d}h_i(p), \quad X\in \sX,
\end{align*}
and consequently, the CDRM under random environment is given by
\begin{align*}
 \rho(X; Z) = \sum^n_{i=1} w_i \int^1_0 \mbox{ES}^{Q_i}_p(X)\mbox{d}h_i(p), \quad X\in \sX,
\end{align*}
which is a $Q$-mixture of ES.

\section{Concluding remarks}

Model uncertainty has been one key issue in risk management and regulation.
In this paper, we take a new perspective to describe the model uncertainty.  More precisely, we use an auxiliary random variable
(called random environment) to describe the model uncertainty.
One advantage of our approach is in its flexibility, because the auxiliary random variable
can describe various contexts including model uncertainty.
We establish a new class of conditional distortion risk measures under random environment. We also axiomatically characterize it
by proposing a set of new axioms.
The coherence and dual representation are further investigated.
To illustrate the proposed framework for risk measures under model uncertainty, we also deduce new risk measures in the presence of background risk.

\vspace{0.2cm}

By checking the proofs, it is clear that most parts of the main results would be still true if we would assume that
the loss random variables are integrable and bounded below. Even more, by certain truncation and approximation approach, one could deal with
more general loss random variables rather than bounded loss random variables. Nevertheless, this is no longer the focus of this paper.
However, from the mathematical point of view, it would be interesting to see it be worked out where the
loss random variables are not necessarily bounded.

\section*{Acknowledgements}

The authors are very grateful to the Editor-in-Chief Professor Frank Riedel and the anonymous referees
for their constructive and valuable comments and suggestions, which led to the present greatly improved version of the manuscript.
The authors are also very grateful to Professor Ruodu Wang for his very helpful discussions and comments on an earlier version of the manuscript.

\section*{Appendix}
\setcounter{equation}{0}
\setcounter{subsection}{0}
\renewcommand{\theequation}{A.\arabic{equation}}
\renewcommand{\thesubsection}{A.\arabic{subsection}}

In this appendix, we provide the proofs of all results of this paper.

\vspace{0.2cm}

\noindent{\bf Proof of Lemma 2.1.}

\vspace{0.2cm}

Since $\phi : \mathbb{R} \rightarrow \mathbb{R}_+$ is $\sB(\mathbb{R})$-measurable, there exists a sequence
of simple functions $\phi_n(z)$ such that $\phi_n (z) \uparrow \phi (z)$ for each $z \in \mathbb{R}.$

\vspace{0.2cm}

For each $n\geq 1$, define a function $g_n : \mathbb{R} \rightarrow \mathbb{R}$ by $g_n(z):= g(z,\phi_n(z))$, $z \in \mathbb{R}.$
Then for every $z \in \mathbb{R},$  by the left-continuity of $g(z,\cdot)$,
\begin{align*}
  \lim_{n\rightarrow \infty} g_n(z)= g(z,\phi(z)).
\end{align*}
Hence, it is sufficient to prove that for each $n \geq 1$, the function $g_n$ is a Borel function on $\mathbb{R}.$

\vspace{0.2cm}

For $n \geq 1$, denote by  $ \mbox{Ran}(\phi_n) := \{a_{n,i} : 1 \leq i \leq k_n\}$ the finite set of values that $\phi_n$ takes.
Hence, for any $b \in \mathbb{R}$,
\begin{align*}
  \left\{ z: g_n(z) \leq b \right\}
   = \underset{1 \leq i \leq k_n}{\cup} \left\{ \left\{ z: g(z, a_{n,i}) \leq b \right\} \cap \left\{ z: \phi_n(z)= a_{n, i} \right\} \right\}
\end{align*}
is a Borel set. Consequently, $g(z,\phi(z))$ is a Borel function on $\mathbb{R}$. Lemma 2.1 is proved.

\vspace{0.2cm}

\noindent{\bf Proof of Proposition 3.1.}

\vspace{0.2cm}

 (1)\ \ Given a random environment $Z \in \sX^\perp $, let the $P_Z-$null set $N_0 \in \sB(\mathbb{R})$ be as in ($\ref{0429add1}$).
Given $z \in N_0^c \cap \mbox{Ran}(Z)$, since $K_Z^*(z, A \cap\{ Z=z \}) := K_Z(z, A)$ for any $A \in \sF$, hence
for any $X \in \sX$, by the definition of $\rho_{_Z}(\cdot; z)$, we have that
\begin{align}\label{add5001}
   \rho_{_Z}(X; z)  = \int X d g_z\circ K_Z(z,\cdot).
\end{align}

\vspace{0.2cm}

By (\ref{add5001}),  Axioms A1 and  A2 are apparent.  Axiom A4 is a direct application of the Monotone Convergence Theorem of Denneberg (1994, Theorem 8.1)
by letting the monotone set function $\mu$ on the $\sigma$-algebra $\sF$ be
$ \mu (A) := g_z \circ K_Z(z, A) = g_z \circ K_Z^*(z, A \cap \{Z = z \}),$ $A \in \sF.$
As for Axiom A3, it is basically a straightforward application of Proposition 5.1(vi) of Denneberg (1994)
or Theorem 4.88 of F\"{o}llmer and Schied (2016) by setting the monotone set function $\mu$ on $\sF$ be $ \mu (A) := g_z \circ K_Z(z, A),$ $A \in \sF.$
In fact, given $X, Y \in \sX$ such that they are local-comonotonic on $\{Z =z\},$ define
\begin{align*}
  X^*(\omega)
       :=\begin{cases}
         X(\omega),\ & \mbox{if}\ \omega \in \{Z =z\}, \\
         -\| X \|, \ & \mbox{if}\ \omega \notin \{Z =z\},
\end{cases}
\quad\quad
 Y^*(\omega)
       :=\begin{cases}
         Y(\omega),\ & \mbox{if}\ \omega \in  \{Z =z\}, \\
         -\| Y \|,\ & \mbox{if}\ \omega \notin \{Z =z\}.
\end{cases}
\end{align*}
Clearly, $X^*$, $Y^*$ $\in$ $\sX$. Moreover, it is not hard to verify that $X^*$ and $Y^*$ are comonotonic on $\Omega$, and that
\begin{align*}
\int X^* d g_z\circ K_Z(z,\cdot) = \int X d g_z\circ K_Z(z,\cdot), \quad
\int Y^* d g_z\circ K_Z(z,\cdot) = \int Y d g_z\circ K_Z(z,\cdot)
\end{align*}
and
\begin{align*}
\int (X^* + Y^*) d g_z\circ K_Z(z,\cdot) = \int (X + Y) d g_z\circ K_Z(z,\cdot).
\end{align*}
Therefore, after setting a monotone set function $\mu$ on $\sF$ be $ \mu (A) := g_z \circ K_Z(z, A),$ $A \in \sF,$
 by applying  Denneberg (1994, Proposition 5.1(vi)) or F\"{o}llmer and Schied (2016, Theorem 4.88) to
the monotone set function $\mu$ on $\sF$ and the comonotonic $X^*$ and $Y^*$,
we obtain that
\begin{align*}
\rho_{_Z}(X + Y; z) = \int (X^* + Y^*) d g_z\circ K_Z(z,\cdot)
                    = \rho_{_Z}(X; z) + \rho_{_Z}(Y; z),
\end{align*}
which is exactly the desired state-wise comonotonic additivity A3.

\vspace{0.2cm}

(2)\ \ Given a random environment $Z \in \sX^\perp $, for each $z \in \mbox{Ran}(Z),$ let the functional $\tau(\cdot; z)$ in Assumption A be the
risk measure $\rho_{_Z}(\cdot; z)$ defined by ($\ref{add3001}$), then by Proposition 3.1(1) we know that Assumption A holds.
Taking ($\ref{add3012}$) and Proposition 3.2(1) into account, then Axioms B1 and B2 are straightforward and clear.

\vspace{0.2cm}

We now show Axiom B3.
Given $Z \in \sX^\perp $, for $X, Y \in \sX$ such that for $P_Z-a.e.$ $z \in \mbox{Ran}(Z)$, $X$ and $Y$ are local-comonotonic on $\{ Z=z \},$
and the two functions $\rho_{_Z}(X; \cdot)$ and $\rho_{_Z}(Y; \cdot)$ are environment-wise comonotonic, then by Proposition 3.2(1),
we know that for every such $P_Z-a.e.$ $z \in \mbox{Ran}(Z)$,
\begin{align*}
    \rho_{_Z}(X+Y; z) = \int (X+Y) d g_z\circ K_Z(z,\cdot)
                   = \rho_{_Z}(X; z) + \rho_{_Z}(Y; z),
\end{align*}
which, together with ($\ref{add3012}$) and Proposition 5.1(vi) of Denneberg (1994) or Theorem 4.88 of F\"{o}llmer and Schied (2016), results to
\begin{align*}
  \rho(X+Y; Z)
      = \int \left[ \rho_{_Z}(X;\cdot)+\rho_{_Z}(Y; \cdot) \right] d h_Z\circ P_Z^*
      = \rho(X;Z) + \rho(Y;Z).
\end{align*}

\vspace{0.2cm}

Finally, we show Axiom B4.  Given $Z \in \sX^\perp $, for  $X, X_n \in \sX,$ $n \geq 1,$
such that $D \leq X_n,$ $ n \geq 1,$ with some constant $D \in \mathbb{R},$ $X_n \uparrow X$ eventually, and $ \rho_{_Z}(X_n; z) \leq \rho_{_Z}(X; z)$
for each $n \geq 1$ and $P_Z-a.e.$ $z \in \mbox{Ran}(Z),$ then for any $\alpha \in \mathbb{R},$
it is not hard to check that for every $\omega \in \Omega,$
\begin{align*}
\lim_{n \rightarrow \infty} 1_{(\alpha, +\infty)}(X_n(\omega)) =  1_{(\alpha, +\infty)}(X(\omega)).
\end{align*}
Hence, by the Lebesgue's Dominated Convergence Theorem, we know that for any above $P_Z-a.e.$ $z \in \mbox{Ran}(Z)$ and any $\alpha \in \mathbb{R},$
\begin{align*}
 \lim_{n \rightarrow \infty} K_Z(z, \{X_n > \alpha \}) = K_Z(z, \{X > \alpha \}),
\end{align*}
and thus,
\begin{align*}
 \lim_{n \rightarrow \infty}g_z \circ K_Z(z, \{X_n > \alpha \}) = g_z \circ K_Z(z, \{X > \alpha \}),
\end{align*}
since $g_z$ is continuous. Therefore,  from (\ref{add5001}) and the Lebesgue's Dominated Convergence Theorem,
it follows that for any above $P_Z-a.e.$ $z \in \mbox{Ran}(Z),$
\begin{align}\label{doublestar1}
 \lim_{n \rightarrow \infty} \rho_{_Z}(X_n; z) = \rho_{_Z}(X; z).
\end{align}

Recall the assumption that $ \rho_{_Z}(X_n; z) \leq \rho_{_Z}(X; z)$ for each $n \geq 1$ and $P_Z-a.e.$ $z\in\mbox{Ran}(Z),$
from (\ref{doublestar1}) it follows that for any  $\beta \in \mathbb{R}$ and any above $P_Z-a.e.$ $z\in\mbox{Ran}(Z)$,
$$
 \lim_{n \rightarrow \infty} 1_{(\beta, +\infty)}(\rho_{_Z}(X_n; z)) =1_{(\beta, +\infty)}(\rho_{_Z}(X; z)),
$$
which, together with the Lebesgue's Dominated Convergence Theorem, implies that for any $\beta\in \mathbb{R}$,
\begin{align}\label{doublestar2}
 \lim_{n \rightarrow \infty} P_Z^*( \{z : \rho_{_Z}(X_n; z) > \beta \} \cap \mbox{Ran}(Z)) = P_Z^*( \{z : \rho_{_Z}(X; z) > \beta \} \cap \mbox{Ran}(Z)).
\end{align}
Notice that for any above $P_Z-a.e.$ $z\in \mbox{Ran}(Z),$  each $n\geq 1$ and any $\beta\in \mathbb{R}$,
\begin{align*}
  P_Z^*  ( \{z : \rho_{_Z}(X_n; z) > \beta \} \cap \mbox{Ran}(Z))
         \leq P_Z^*(\{ z: \rho_{_Z}(X; z) > \beta \} \cap \mbox{Ran}(Z)),
\end{align*}
which, as well as  (\ref{doublestar2}) and the left-continuity of $h_Z$, yields that for any  $\beta\in \mathbb{R}$,
\begin{align}\label{1115add1}
 \lim_{n \rightarrow \infty}
         & h_Z \circ  P_Z^*( \{z : \rho_{_Z}(X_n; z) > \beta \} \cap \mbox{Ran}(Z)) \nonumber\\
         & = h_Z \circ  P_Z^* ( \{z : \rho_{_Z}(X; z) > \beta \} \cap \mbox{Ran}(Z)).
\end{align}

\vspace{0.2cm}

Keeping ($\ref{add3012}$) in mind, by (\ref{1115add1}) and the Lebesgue's Dominated Convergence Theorem,
we know that
$
\lim_{n \rightarrow \infty} \rho(X_n; Z) = \rho (X; Z),
$
which is just the desired assertion. Proposition 3.1 is proved.

\vspace{0.2cm}

\noindent{\bf Proof of Proposition 3.2.}

\vspace{0.2cm}

Basically speaking, Proposition 3.2 is an application of Corollary 13.3 of Denneberg (1994), or Schmeidler's Representation Theorem
(for example, see Theorem 11.2 of Denneberg (1994)). However, since there will be various $P_Z-$null sets involved in the sequel
(for instance, see the proof of Proposition 3.3 below), and there is also certain equivalent relation needed to be pointed out
(see (\ref{0430add1}) below), we sketch the proof here.

\vspace{0.2cm}

Fix arbitrarily a random environment $Z \in \sX^\perp $. Let the $P_Z-$null set $N_0 \in \sB(\mathbb{R})$ be as in ($\ref{0429add1}$),
and denote by $B_0 \in \sB(\mathbb{R})$ the $P_Z-$null set with $N_0 \subseteq B_0$ such that for every $z \in B_0^c \cap \mbox{Ran}(Z),$
the normalized risk measure $\rho_{_Z}(\cdot; z)$ satisfies Axioms A2 and A3. Note first that given a state $z \in B_0^c \cap \mbox{Ran}(Z),$
for any $X \in \sX$, we have that
\begin{align}\label{0430add1}
\rho_{_Z}(X; z) = \rho_{_Z}(X1_{ \{Z=z\} }; z).
\end{align}
Indeed, (\ref{0430add1}) is due to Axiom A2, because $X$ and $X1_{ \{Z=z\} }$ are equal on $\{Z= z\}.$

\vspace{0.2cm}

Given arbitrarily a state $z \in B_0^c \cap \mbox{Ran}(Z),$ we first claim that the normalized risk measure $\rho_{_Z}(\cdot; z)$ is monotone,
that is $\rho_{_Z}(X; z) \leq \rho_{_Z}(Y; z)$ for any $X, Y \in \sX$ with $X(\omega) \leq Y(\omega)$
for every $\omega \in \Omega.$
In fact, given any $X, Y \in \sX$ with $X(\omega) \leq Y(\omega)$ for every $\omega \in \Omega,$
then $X(\omega) \leq Y(\omega)$ for every $\omega \in \{Z = z\}$. Hence, from Axiom A2, it follows that
$\rho_{_Z}(X; z) \leq  \rho_{_Z}(Y; z).$

\vspace{0.2cm}

Next, we claim that the normalized risk measure $\rho_{_Z}(\cdot; z)$ is comonotonic additive,
that is
\begin{align*}
\rho_{_Z}(X + Y; z) = \rho_{_Z}(X; z) + \rho_{_Z}(Y; z)
\end{align*}
for any $X, Y \in \sX$ so that $X$ and $Y$ are comonotonic on $\Omega.$
In fact, given any $X, Y \in \sX$ so that $X$ and $Y$ are comonotonic on $\Omega,$
then clearly, $X$ and $Y$ are local-comonotonic on $\{Z = z\}$. Hence, from Axiom A3, it follows that
\begin{align*}
\rho_{_Z}(X + Y; z) = \rho_{_Z}(X; z) + \rho_{_Z}(Y; z).
\end{align*}
Consequently, by the Schmeidler's Representation Theorem (for instance, see Theorem 11.2 of Denneberg (1994))
or Corollary 13.3 of Denneberg (1994), we know that there is a monotone and normalized set function $\gamma_z$ on $\sF$
such that for any $X \in \sX_+,$
\begin{align*}
\rho_{_Z}(X ; z) = \int_0^{\infty} \gamma_z( \{ X > \alpha \} ) d\alpha.
\end{align*}
Particularly, for every $A \in \sF,$ $\gamma_z(A) := \rho_{_Z}(1_A; z).$ Taking (\ref{0430add1}) into account,
we also know that for any $X \in \sX_+,$
\begin{align*}
\rho_{_Z}(X ; z) = \rho_{_Z}(X1_{ \{Z = z\} } ; z) = \int_0^{\infty} \gamma_z( \{ X > \alpha \} \cap \{ Z = z \} ) d\alpha,
\end{align*}
and that for every $A \in \sF,$
\begin{align*}
\gamma_z(A) := \rho_{_Z}(1_A; z) = \rho_{_Z}(1_{ A \cap \{Z = z\} }; z).
\end{align*}
Proposition 3.2 is proved.

\vspace{0.2cm}

\noindent{\bf Proof of Proposition 3.3.}

\vspace{0.2cm}

The key points of the proof are basically the same as ones of proof of Theorem 3 of Wang et al. (1997) by replacing the probability measure
there with probability measure $K_Z(z,\cdot)$ provided with a random environment $Z \in \sX^\perp $ and certain $z \in \mbox{Ran}(Z)$.
Nevertheless, since there is a little distinction between the assumptions employed as pointed out in Remark 3.3, while
we also concern the continuity issue of the distortion function involved, we would like to provide an alternative proof to construct
the desired distortion function.

\vspace{0.2cm}

Let a random environment $Z \in \sX^\perp $ be fixed, and $U$ a U[0, 1] random variable on $(\Omega, \sF, P)$ independent of $Z$.
Let the $P_Z-$null set $N_0$ be as in ($\ref{0429add1}$), and denote by $B_0 \in \sB(\mathbb{R})$
the $P_Z-$null set with $N_0 \subseteq B_0$ such that for every $z \in B_0^c \cap \mbox{Ran}(Z),$
the normalized risk measure $\rho_{_Z}(\cdot; z)$ satisfies Axioms A1-A4.
From ($\ref{0415add1}$), we know that there is a $P_Z-$null set $N \in \sB(\mathbb{R})$
such that for any $z \in N^c \cap \mbox{Ran}(Z)$, $U$ is also uniformly distributed on $[0, 1]$ under probability measure  $K_Z(z, \cdot).$
Furthermore, we can choose the $P_Z-$null set $N$ large enough such that $B_0 \subseteq N$.
Therefore, for every $z \in N^c \cap \mbox{Ran}(Z)$, from Proposition 3.2 it follows that there exists a monotone and normalized set function
$\gamma_z$ on $\sF$ depending on $z$ and uniquely determined by $\rho_{_Z}(\cdot;z)$ such that for any $X \in \sX_+$,
\begin{align}\label{4002}
  \rho_{_Z}(X; z)= \int_0^{\infty} \gamma_z(\{ X > \alpha \})d\alpha.
\end{align}
Note that for every  $A \in \sF$, $\gamma_z(A) := \rho_{_Z}(1_A; z)$.

\vspace{0.2cm}

For each $z \in N^c\cap\mbox{Ran}(Z)$, define a function $g_z: [0,1] \rightarrow \mathbb{R}_+$ by
\begin{align}\label{add5002}
  g_z(u):= \rho_{_Z} \left( 1_{\{ U > 1-u \}}; z \right).
\end{align}
Axiom A4 yields that $g_z$ is left-continuous.
Moreover, by the normalization of $\rho_{_Z}(\cdot; z)$ and Axioms A1 and A2, it is not hard to verify that $g_z$ has the following
three properties:
(1)\ $g_z (0)= 0$,  (2)\ $g_z (1)= 1$, and (3) $g_z (u) \leq g_z(v)$ for any $0\leq u \leq v \leq 1$.
Thus, the function $g_z$ is a left-continuous distortion function.

\vspace{0.2cm}

Given arbitrarily a state $ z\in N^c \cap \mbox{Ran}(Z),$ for any $X \in \sX_+$ and any $t \geq 0$,
note that
$$
K_Z(z, \{ X > t \}) = 1- [1-K_Z(z,\{  X > t \})] = K_Z(z, \{U > 1-K_Z(z,\{  X > t \})),
$$
hence, the Bernoulli random variables $1_{\{  X > t \}}$ and $1_{\{U > 1-K_Z(z,\{ X > t \})\}}$ have the same probability distribution
with respect to $K_Z(z, \cdot)$.
Therefore, by the definitions of $\gamma_z$ and $g_z$ and Axiom A1, we have that
\begin{align*}
  \gamma_z( \{ X > t \})
       = \rho_{_Z}(1_{ \{ X > t \} }; z)
       = \rho_{_Z}(1_{\{U > 1-K_Z(z,\{  X > t \})\}}; z)
       = g_z(K_Z(z,\{ X > t \})),
\end{align*}
which, together with $(\ref{4002}),$  yields
\begin{align*}
  \rho_{_Z}(X; z) = \int_0^{\infty} \gamma_z ( \{ X > \alpha \} ) d\alpha
               = \int_0^{\infty} g_z \circ K_Z(z,\{ X > \alpha \}) d\alpha,
\end{align*}
which proves the proposition. Proposition 3.3 is proved.

\vspace{0.2cm}

\noindent{\bf Proof of Proposition 3.4.}

\vspace{0.2cm}

By the same approach as that in the Appendix of Wang et al. (1997), we can extend Proposition 3.3
to real-valued random variables. We sketch the proof here.

\vspace{0.2cm}

Let a random environment $Z \in \sX^\perp $ be fixed. Given any $X \in \sX, $ then for any $m <0$, $X\vee m- m \in \sX_+.$
By Proposition 3.3 we know that there exists a $P_Z-$null set $N\in \sB(\mathbb{R})$ such that for any $ z \in N^c \cap \mbox{Ran}(Z), $
$$
\rho_{_Z}(X\vee m - m; z)= \int_0^{\infty} g_z \circ K_Z(z,\{ X\vee m- m > \alpha \})d\alpha,
$$
where the left-continuous distortion function $g_z$ is defined by (\ref{add5002}).

\vspace{0.2cm}

Keeping in mind the fact that a constant (regarded as a degenerate random variable) and any random variable are local-comonotonic on
$\{Z = z \}$, by Axiom A3 we know that
\begin{align*}
  \rho_{_Z}(X\vee m- m;z)= \rho_{_Z}(X\vee m; z) + \rho_{_Z}(-m; z) = \rho_{_Z}(X\vee m; z) + (-m),
\end{align*}
in which that $\rho_{_Z}(-m; z) = -m$ has been used, which is due to the positive homogeneity and normalization of $\rho_{_Z}(\cdot; z)$.
Thus, by change-of-variable,
\begin{align}\label{doublestar3}
 \rho_{_Z}(X \vee m; z)
    & =  \int_0^{\infty} g_z \circ K_Z(z,\{ X\vee m - m > \alpha \})d\alpha - (-m) \nonumber \\
    & =  \int_m^0 (g_z \circ K_Z(z,\{ X> \alpha \}) -1)d\alpha + \int_0^{\infty} g_z \circ K_Z(z,\{ X > \alpha \})d\alpha.
\end{align}

Since $X$ is bounded,  we can choose $ m $ small enough, for example less than $ - \|X\|, $ so that $ \rho_{_Z}(X \vee m; z) = \rho_{_Z}(X; z)$
and
\begin{align*}
\int_m^0 & (g_z \circ K_Z(z,\{ X> \alpha \}) -1)d\alpha + \int_0^{\infty} g_z \circ K_Z(z,\{ X > \alpha \})d\alpha \\
         & = \int_{-\infty}^0  (g_z \circ K_Z(z,\{ X> \alpha \}) -1)d\alpha + \int_0^{\infty} g_z \circ K_Z(z,\{ X > \alpha \})d\alpha.
\end{align*}
Consequently, from (\ref{doublestar3}) it follows that
\begin{align*}
  \rho_{_Z}(X; z)
     = \int_{-\infty}^0 \left(g_z \circ K_Z(z,\{ X > \alpha \}) -1\right)d\alpha
          + \int_0^{\infty} g_z \circ K_Z(z,\{ X > \alpha \})d\alpha,
\end{align*}
which shows the desired assertion. Proposition 3.4 is proved.

\vspace{0.2cm}

\noindent{\bf Proof of Theorem 3.1.}

\vspace{0.2cm}

Let a random environment $Z \in \sX^\perp $ be given, and let the family $\{ \tau_Z(\cdot; z) ;$ $z \in \mbox{Ran}(Z) \}$
of functionals be as in Assumption A.
Applying Propositions 3.3 and 3.4 to the functionals $\{ \tau_Z(\cdot; z) ;$ $z \in \mbox{Ran}(Z) \}$ implies that there is
a $P_Z-$null set $N\in \sB(\mathbb{R})$, so that for each $z \in N^c \cap \mbox{Ran}(Z),$
there exists a left-continuous distortion function $g_z$ defined by (\ref{add5002}) such that for any $X \in \sX_+$,
\begin{align}\label{5001}
\tau_Z(X;z) = \int_0^{\infty} g_z \circ K_Z(z,\{ X > \alpha \})d\alpha,
\end{align}
and that for any $X \in \sX$,
\begin{align}\label{add040301}
\tau_Z(X;z) = \int_{-\infty}^0 \left[ g_z \circ K_Z(z,\{ X > \alpha \}) - 1 \right] d\alpha
                 + \int_0^{\infty} g_z \circ K_Z(z,\{ X > \alpha \})d\alpha.
\end{align}

\vspace{0.2cm}

To show the theorem, it is sufficient for us to show that there exists a monotone and normalized set function $\gamma_Z$ on $\sB(\mbox{Ran}(Z))$
depending on $Z$ such that for any $X \in \sX_+$,
\begin{align}\label{5017}
\rho(X;Z) = \int_0^{\infty} \gamma_Z\left\{ z \in N^c\cap\mbox{Ran}(Z) : \tau_Z(X;z) >\beta \right\}d\beta.
\end{align}
Note that in (\ref{5017}), for the simplicity of notations, we have dropped the parentheses after the set function $\gamma_Z$,
which should bracket the relevant measurable set, and this should not prevent one from right understanding.
Therefore, we will also keep dropping this kinds of parentheses accordingly in the subsequent proofs.

\vspace{0.2cm}

To this end, we first define two set functions $\alpha_Z$, $\beta_Z$ : $\sB(\mbox{Ran}(Z)) \rightarrow \mathbb{R}_+$, through
\begin{align*}
  & \alpha_Z(B^*) := \sup\{ \rho(X; Z): X \in \sX, \tau_Z(X;z) \leq 1_{B^*}(z)\ \mbox{for any } z \in N^c \cap \mbox{Ran}(Z) \}, \\
  & \beta_Z(B^*) := \inf\{ \rho(Y; Z): Y \in \sX,  1_{B^*}(z) \leq \tau_Z(Y; z)  \ \mbox{for any } z \in N^c \cap \mbox{Ran}(Z) \},
\end{align*}
$B^* \in \sB(\mbox{Ran}(Z))$. Indeed, it is not hard to verify that $\alpha_Z(\emptyset) = \beta_Z(\emptyset) = 0$, and that $\alpha_Z$ and $\beta_Z$
are monotone. By Axiom B2, $\alpha_Z \leq \beta_Z.$ Thus, $\alpha_Z$ and $\beta_Z$ are normalized as well.

\vspace{0.2cm}

Let $\gamma_Z: \sB(\mbox{Ran}(Z)) \rightarrow \mathbb{R}_+$ be a monotone and normalized set function such that
\begin{align*}
\alpha_Z \leq \gamma_Z \leq \beta_Z,
\end{align*}
for example, we can choose $\gamma_Z = \alpha_Z$.

\vspace{0.2cm}

Given $X \in \sX_+$, for any positive integer $n$ and each $i \in \{ 1, \cdots, n\cdot 2^{n}-1 \}$,
define two random variables $ X_{i,1}^*, X_{i,2}^* : (\Omega, \sF) \rightarrow \mathbb{R} $ by
\begin{align*}
  X_{i,1}^*(\omega)
  :=\begin{cases}
  \frac{1}{2^n}, \quad\quad\quad & \mbox{if}\quad \tau_Z(X;Z(\omega))> \frac{i+1}{2^n}, \\
  X(\omega)-\frac{i}{2^n}, \quad\quad\quad & \mbox{if}\quad \frac{i}{2^n}< \tau_Z(X;Z(\omega))\leq \frac{i+1}{2^n}, \\
  0, \quad\quad\quad & \mbox{if}\quad 0 < \tau_Z(X;Z(\omega))\leq \frac{i}{2^n}, \\
  \frac{1}{n\cdot 2^{n}-1} \cdot X(\omega), & \mbox{if}\quad \tau_Z(X;Z(\omega)) = 0,
  \end{cases}
\end{align*}
and
\begin{align*}
  X_{i,2}^*(\omega)
  :=\begin{cases}
  \frac{1}{2^n}, \quad\quad\quad &\mbox{if}\quad \tau_Z(X;Z(\omega))> \frac{i}{2^n}, \\
  X(\omega)-\frac{i-1}{2^n}, \quad\quad\quad & \mbox{if}\quad \frac{i-1}{2^n}< \tau_Z(X;Z(\omega))\leq \frac{i}{2^n}, \\
  0, \quad\quad\quad & \mbox{if}\quad 0 < \tau_Z(X;Z(\omega))\leq \frac{i-1}{2^n}, \\
  \frac{1}{n\cdot 2^{n}-1} \cdot X(\omega), & \mbox{if}\quad \tau_Z(X;Z(\omega)) = 0.
  \end{cases}
\end{align*}
In fact, the $\sF$-measurability of $X_{i,1}^*$ and $X_{i,2}^*$ follows from the facts that $\{ \tau_Z(X; z); z\in \mbox{Ran}(Z) \}$
is regular on $\mbox{Ran}(Z)$, and that $Z$ is $\sF/\sB(\mbox{Ran}(Z))$-measurable. Particularly, we note that for any
$t \in \mathbb{R}_+$, $\left\{ \omega : \tau_Z(X;Z(\omega))> t \right\}$
is a $\sF$-measurable set. Moreover, it is not hard to verify that for each $i \in \{ 1, \cdots, n\cdot 2^{n}-1 \}$, $X_{i,1}^*$ and $X_{i,2}^*$ are local-comonotonic on $\{ Z=z \}$ for each $z \in \mbox{Ran}(Z)$.

\vspace{0.2cm}

Next, we will show that (\ref{5017}) holds for $\gamma_Z$ defined as above, and thus complete the proof of the theorem. We divide the proof
into three steps.

\vspace{0.2cm}

Step one. We conclude that for any positive integer $n \geq 1$ and each $i \in \{ 1, \cdots, n\cdot 2^{n}-1 \}$,
\begin{align}\label{5018}
 \rho \left( \sum_{i=1}^{n\cdot 2^n-1} X_{i,1}^*;Z \right)
    & \leq \frac{1}{2^n}\sum_{i=1}^{n\cdot 2^n-1} \gamma_Z\left\{ z \in N^c \cap \mbox{Ran}(Z) : \tau_Z(X;z)> \frac{i}{2^n} \right\} \nonumber\\
    & \leq \rho\left( \sum_{i=1}^{n\cdot 2^n-1} X_{i,2}^*;Z \right).
\end{align}

\vspace{0.2cm}

To this end, we first show that for any  $z \in N^c \cap \mbox{Ran}(Z)$ and each $i \in \{ 1, \cdots, n\cdot 2^{n}-1 \}$,
\begin{align}\label{5032}
  \tau_Z(2^n \cdot X_{i,1}^*; z) \leq 1_{\left\{ \tau_Z(X;\cdot)>\frac{i}{2^n} \right\}}(z) \leq \tau_Z(2^n \cdot X_{i,2}^*; z).
\end{align}

\vspace{0.2cm}

Given arbitrarily $z \in N^c \cap \mbox{Ran}(Z)$, we now calculate $ \tau_Z(X_{i,1}^*;z)$. Note that $ 0 \leq X_{i,1}^* + \frac{i}{2^n}$, and that
$X_{i,1}^* $ and $ \frac{i}{2^n}$ are local-comonotonic on $\{ Z =z \}$, from Axiom A3 and Remark 3.1(ii),
it follows that
\begin{align*}
 \tau_Z \left( X_{i,1}^* + \frac{i}{2^n}; z \right)
   = \tau_Z \left( X_{i,1}^*; z \right) + \tau_Z \left( \frac{i}{2^n}; z \right)
   = \tau_Z(X_{i,1}^*;z) + \frac{i}{2^n},
\end{align*}
which yields that
\begin{align}\label{add5021}
  \tau_Z( X_{i,1}^*;z) = \tau_Z \left( X_{i,1}^* + \frac{i}{2^n}; z \right)  - \frac{i}{2^n}.
\end{align}

Based on (\ref{5001}) and (\ref{add5021}), we calculate $\tau_Z( X_{i,1}^*;z).$
Since $\tau_Z(X;z)$ is non-negative,
there are four possibilities for the value of $\tau_Z(X; z)$.

Case one: $\tau_Z(X;z) > \frac{i+1}{2^n}$.
In this case, by the definition of $X_{i,1}^*$, we know that $X_{i,1}^*(\omega) = \frac{1}{2^n}$ for $\omega \in \{ Z=z \}$.
Hence, by (\ref{5001}) and  (\ref{add5021}),
\begin{align}\label{5002}
 \tau_Z(X_{i,1}^*;z)
   & = \int_0^{\infty} g_z \circ K_Z\left(z,\left\{ \omega : X_{i,1}^*(\omega)+ \frac{i}{2^n} > \alpha \right\}\right)d\alpha - \frac{i}{2^n} \nonumber\\
   & = \int_0^{\infty} g_z \circ K_Z^*\left(z,\left\{ \omega : X_{i,1}^*(\omega)+ \frac{i}{2^n} > \alpha, Z(\omega)=z \right\}\right)d\alpha
       - \frac{i}{2^n} \nonumber\\
   & = \int_0^{\infty} g_z \circ K_Z^*\left(z,\left\{ \omega : \frac{1}{2^n}+ \frac{i}{2^n} > \alpha, Z(\omega)=z \right\}\right)d\alpha
       - \frac{i}{2^n} \nonumber\\
   & = \int_0^{\infty} g_z \circ K_Z\left(z,\left\{ \omega : \frac{1}{2^n}+ \frac{i}{2^n} > \alpha \right\}\right)d\alpha - \frac{i}{2^n} \nonumber\\
   & = \frac{1}{2^n}.
\end{align}

By the same argument similar to above Case one, we can calculate $\tau_Z( X_{i,1}^*;z)$ for the other three cases as follows.

Case two: $\frac{i}{2^n} < \tau_Z(X;z) \leq \frac{i+1}{2^n}$. Then, by (\ref{5001}) and (\ref{add5021}),
\begin{align}\label{5003}
 \tau_Z(X_{i,1}^*;z)
    = \tau_Z\left(X_{i,1}^*+ \frac{i}{2^n};z\right) - \frac{i}{2^n}
    = \tau_Z(X;z) - \frac{i}{2^n}.
\end{align}

Case three: $0 < \tau_Z(X;z) \leq \frac{i}{2^n}$. Then, by (\ref{5001}) and  (\ref{add5021}),
\begin{align}\label{5004}
 \tau_Z(X_{i,1}^*;z)
   = \tau_Z\left(X_{i,1}^*+ \frac{i}{2^n};z\right) - \frac{i}{2^n}
   = 0.
\end{align}

Case four: $\tau_Z(X;z) = 0$. Then, by (\ref{5001}) and (\ref{add5021}),
\begin{align}\label{5033}
 \tau_Z(X_{i,1}^*;z)
     = \tau_Z\left(X_{i,1}^*+ \frac{i}{2^n};z\right) - \frac{i}{2^n}
     = 0.
\end{align}

\vspace{0.2cm}

In summary, for any $z \in N^c \cap \mbox{Ran}(Z)$ and each $i \in \{ 1, \cdots, n\cdot 2^{n}-1 \}$,
\begin{align}\label{5019}
  \tau_Z(X_{i,1}^*;z) \leq \frac{1}{2^n} \cdot 1_{ \left\{ \tau_Z(X; \cdot) > \frac{i}{2^n} \right\}}(z).
\end{align}
By a similar argumentation as above, we can also steadily show that for any $z \in N^c \cap \mbox{Ran}(Z)$ and
each $i \in \{ 1, \cdots, n\cdot 2^{n}-1 \}$,
\begin{align*}
  \tau_Z( X_{i,2}^*;z)
      = \begin{cases}
            \frac{1}{2^n}, \quad\quad\quad & \mbox{if}\quad \tau_Z(X;z)> \frac{i}{2^n}, \\
            \tau_Z(X;z)-\frac{i-1}{2^n}, \quad\quad\quad & \mbox{if}\quad \frac{i-1}{2^n}< \tau_Z(X;z)\leq \frac{i}{2^n}, \\
            0, \quad\quad\quad & \mbox{if}\quad 0 \leq \tau_Z(X;z)\leq \frac{i-1}{2^n}.
       \end{cases}
\end{align*}
Hence
\begin{align}\label{5020}
  \frac{1}{2^n} \cdot 1_{ \left\{ \tau_Z(X; \cdot) > \frac{i}{2^n} \right\}}(z) \leq \tau_Z(X_{i,2}^*;z).
\end{align}
Thus, recalling that $\tau_Z(\cdot;z)$ is positive homogeneous (see Remark 3.1(ii)),  (\ref{5019}) and (\ref{5020}) together yield
\begin{align*}
  \tau_Z(2^n \cdot X_{i,1}^*; z) \leq 1_{\left\{ \tau_Z(X;\cdot)>\frac{i}{2^n} \right\}}(z) \leq \tau_Z(2^n \cdot X_{i,2}^*; z),
\end{align*}
which is just (\ref{5032}).

\vspace{0.2cm}

Note that $\left\{z \in N^c \cap \mbox{Ran}(Z) : \tau_Z(X; z) > \frac{i}{2^n} \right\} \in \sB(\mbox{Ran}(Z))$, from (\ref{5032}) and
the definitions of $\alpha_Z, \gamma_Z$ and $\beta_Z,$ it follows that
\begin{align*}
  \rho(2^n \cdot X_{i,1}^*;Z)
    & \leq  \alpha_Z{\left\{ z \in N^c \cap \mbox{Ran}(Z) : \tau_Z(X;z)> \frac{i}{2^n} \right\}} \nonumber\\
    & \leq  \gamma_Z{\left\{ z \in N^c \cap \mbox{Ran}(Z) : \tau_Z(X;z)> \frac{i}{2^n} \right\}} \nonumber\\
    & \leq  \beta_Z{\left\{ z \in N^c \cap \mbox{Ran}(Z) : \tau_Z(X;z)> \frac{i}{2^n} \right\}} \nonumber\\
    & \leq  \rho(2^n \cdot X_{i,2}^*;Z).
\end{align*}
Thus, by the positive homogeneity of $\rho(\cdot; Z)$ (see Remark 3.1(ii)),
\begin{align*}
 2^n \cdot  \rho(X_{i,1}^*;Z)
    \leq  \gamma_Z{\left\{ z \in N^c \cap \mbox{Ran}(Z) : \tau_Z(X;z)> \frac{i}{2^n} \right\}}
    \leq  2^n \cdot \rho(X_{i,2}^*;Z),
\end{align*}
which implies that for any $n \geq 1$,
\begin{align}\label{5034}
  \sum_{i=1}^{n\cdot 2^n -1} \rho(X_{i,1}^*; Z)
    & \leq \frac{1}{2^n} \sum_{i=1}^{n\cdot 2^n -1} \gamma_Z{\left\{ z \in N^c \cap \mbox{Ran}(Z) : \tau_Z(X;z)> \frac{i}{2^n} \right\}} \nonumber\\
    & \leq \sum_{i=1}^{n\cdot 2^n -1} \rho(X_{i,2}^*; Z).
\end{align}

\vspace{0.2cm}

Observing (\ref{5034}), to show (\ref{5018}), it is sufficient for us to show
\begin{align}\label{5035}
  \sum_{i=1}^{n\cdot 2^n -1} \rho(X_{i,1}^*; Z) = \rho\left(\sum_{i=1}^{n\cdot 2^n -1} X_{i,1}^*; Z\right)
\end{align}
and
\begin{align}\label{5036}
  \sum_{i=1}^{n\cdot 2^n -1} \rho(X_{i,2}^*; Z) = \rho\left(\sum_{i=1}^{n\cdot 2^n -1} X_{i,2}^*; Z\right).
\end{align}

\vspace{0.2cm}

Next, we claim that for any fixed $m \in \{ 1, \cdots, n\cdot 2^{n}-2 \}$, the two functions $\tau_Z \left( \sum_{i=1}^{m} X_{i,1}^*; \cdot \right)$
and  $\tau_Z( X_{m+1,1}^*; \cdot)$ are environment-wise comonotonic.  In fact, by the definition of $X_{i,1}^*$,
\begin{align}\label{5014}
  \sum_{i=1}^{m} X_{i,1}^*(\omega)
  =\begin{cases}
  \frac{m}{2^n}, \quad\quad\quad & \mbox{if}\quad \tau_Z(X;Z(\omega))> \frac{m+1}{2^n}, \\
  X(\omega)-\frac{1}{2^n}, \quad\quad\quad & \mbox{if}\quad \frac{1}{2^n}< \tau_Z(X;Z(\omega))\leq \frac{m+1}{2^n}, \\
  0, \quad\quad\quad & \mbox{if}\quad 0 < \tau_Z(X;Z(\omega))\leq \frac{1}{2^n}, \\
  \frac{m}{n\cdot 2^{n}-1} \cdot X(\omega), & \mbox{if}\quad \tau_Z(X;Z(\omega)) = 0,
  \end{cases}
\end{align}
and
\begin{align}\label{5015}
  X_{m+1,1}^*(\omega)
  =\begin{cases}
  \frac{1}{2^n}, \quad\quad\quad & \mbox{if}\quad \tau_Z(X;Z(\omega))> \frac{m+2}{2^n}, \\
  X(\omega)-\frac{m+1}{2^n}, \quad\quad\quad & \mbox{if}\quad \frac{m+1}{2^n}< \tau_Z(X;Z(\omega))\leq \frac{m+2}{2^n}, \\
  0, \quad\quad\quad & \mbox{if}\quad \tau_Z(X;Z(\omega))\leq \frac{m+1}{2^n}, \\
  \frac{1}{n\cdot 2^{n}-1} \cdot X(\omega), & \mbox{if}\quad \tau_Z(X;Z(\omega)) = 0.
  \end{cases}
\end{align}

\vspace{0.2cm}

By an elementary calculation similar to that of $\tau_Z(X_{i,1}^*;z)$, i.e. (\ref{5002})--(\ref{5033}),
we can obtain that for any $z \in N^c \cap \mbox{Ran}(Z)$,
\begin{align}\label{5005}
  \tau_Z\left(\sum_{i=1}^{m} X_{i,1}^*;z\right)
  =\begin{cases}
  \frac{m}{2^n}, \quad\quad\quad & \mbox{if}\quad \tau_Z(X;z)> \frac{m+1}{2^n}, \\
  \tau_Z(X;z)-\frac{1}{2^n}, \quad\quad\quad & \mbox{if}\quad \frac{1}{2^n}< \tau_Z(X;z)\leq \frac{m+1}{2^n}, \\
  0, \quad\quad\quad & \mbox{if}\quad 0 \leq \tau_Z(X;z)\leq \frac{1}{2^n},
  \end{cases}
\end{align}
and
\begin{align}\label{5023}
  \tau_Z( X_{m+1,1}^*;z)
  =\begin{cases}
  \frac{1}{2^n}, \quad\quad\quad & \mbox{if}\quad \tau_Z(X;z)> \frac{m+2}{2^n}, \\
  \tau_Z(X;z)-\frac{m+1}{2^n}, \quad\quad\quad & \mbox{if}\quad \frac{m+1}{2^n}< \tau_Z(X;z)\leq \frac{m+2}{2^n}, \\
  0, \quad\quad\quad & \mbox{if}\quad 0 \leq \tau_Z(X;z)\leq \frac{m+1}{2^n}.
  \end{cases}
\end{align}
Hence, we can steadily verify that the two functions
$\tau_Z\left(\sum_{i=1}^{m} X_{i,1}^*; \cdot\right)$ and $\tau_Z( X_{m+1,1}^*; \cdot)$ are environment-wise comonotonic.
Meanwhile, from (\ref{5014}) and (\ref{5015}), it is clear that given any $z \in N^c \cap \mbox{Ran}(Z)$,
for $m \in \{ 1, \cdots, n\cdot 2^n - 2 \}$, $\sum_{i=1}^{m} X_{i,1}^*$ and
$X_{m+1,1}^*$ are local-comonotonic on $\{ Z = z \}$.
Therefore, by Axiom B3,
\begin{align*}
        \sum_{i=1}^{n\cdot 2^n -1} \rho(X_{i,1}^*; Z) = \rho\left(\sum_{i=1}^{n\cdot 2^n -1} X_{i,1}^*; Z\right),
\end{align*}
which is just (\ref{5035}).

\vspace{0.2cm}

Similarly, we have that for any $z \in N^c \cap \mbox{Ran}(Z)$ and each $m \in \{ 1, \cdots, n\cdot 2^{n}-2 \}$,
\begin{align}\label{5025}
  \sum_{i=1}^{m} X_{i,2}^*(\omega)
    = \begin{cases}
         \frac{m}{2^n}, \quad\quad\quad & \mbox{if}\quad \tau_Z(X;Z(\omega)) > \frac{m}{2^n}, \\
         X(\omega), \quad\quad\quad & \mbox{if}\quad 0 < \tau_Z(X;Z(\omega))\leq \frac{m}{2^n}, \\
        \frac{m}{n\cdot 2^{n}-1} \cdot X(\omega), & \mbox{if}\quad \tau_Z(X;Z(\omega)) = 0,
    \end{cases}
\end{align}

\begin{align}\label{5026}
  X_{m+1,2}^*(\omega)
    = \begin{cases}
         \frac{1}{2^n}, \quad\quad\quad & \mbox{if}\quad \tau_Z(X;Z(\omega))> \frac{m+1}{2^n}, \\
         X(\omega)-\frac{m}{2^n}, \quad\quad\quad & \mbox{if}\quad \frac{m}{2^n}< \tau_Z(X;Z(\omega))\leq \frac{m+1}{2^n}, \\
         0, \quad\quad\quad & \mbox{if}\quad 0 < \tau_Z(X;Z(\omega))\leq \frac{m}{2^n}, \\
          \frac{1}{n\cdot 2^{n}-1} \cdot X(\omega), & \mbox{if}\quad \tau_Z(X;Z(\omega)) = 0,
    \end{cases}
\end{align}

\begin{align}\label{5027}
  \tau_Z\left(\sum_{i=1}^{m} X_{i,2}^*;z\right)
     = \begin{cases}
             \frac{m}{2^n}, \quad\quad\quad & \mbox{if}\quad \tau_Z(X;z)> \frac{m}{2^n}, \\
             \tau_Z(X;z), \quad\quad\quad & \mbox{if}\quad 0 \leq \tau_Z(X;z)\leq \frac{m}{2^n},
       \end{cases}
\end{align}
and
\begin{align}\label{5028}
  \tau_Z( X_{m+1,2}^*;z)
      = \begin{cases}
            \frac{1}{2^n}, \quad\quad\quad & \mbox{if}\quad \tau_Z(X;z)> \frac{m+1}{2^n}, \\
            \tau_Z(X;z)-\frac{m}{2^n}, \quad\quad\quad & \mbox{if}\quad \frac{m}{2^n}< \tau_Z(X;z)\leq \frac{m+1}{2^n}, \\
            0, \quad\quad\quad & \mbox{if}\quad 0 \leq \tau_Z(X;z)\leq \frac{m}{2^n}.
       \end{cases}
\end{align}
Hence,  we can steadily verify that the two functions $\tau_Z\left(\sum_{i=1}^{m} X_{i,2}^*; \cdot\right)$ and $\tau_Z( X_{m+1,2}^*; \cdot)$
are environment-wise comonotonic, and that
 $\sum_{i=1}^{m} X_{i,2}^*$ and $X_{m+1,2}^*$ are local-comonotonic on $\{ Z = z \}$ for each $z \in N^c \cap \mbox{Ran}(Z)$.
Therefore, by Axiom B3,
\begin{align*}
        \sum_{i=1}^{n\cdot 2^n -1} \rho(X_{i,2}^*; Z) = \rho\left(\sum_{i=1}^{n\cdot 2^n -1} X_{i,2}^*; Z\right),
\end{align*}
which is just (\ref{5036}).
Consequently, (\ref{5018}) follows from (\ref{5034})--(\ref{5036}).

\vspace{0.2cm}

Step two. We continue to conclude that
\begin{align}\label{5037}
  \underset{n \rightarrow \infty}{\lim} \rho\left(\sum_{i=1}^{n\cdot 2^n -1} X_{i,1}^*; Z\right)
  = \underset{n \rightarrow \infty}{\lim} \rho\left(\sum_{i=1}^{n\cdot 2^n -1} X_{i,2}^*; Z\right)
  = \rho(X;Z).
\end{align}

To this end, we first show that
\begin{align}\label{5007}
\lim_{n\rightarrow \infty} \rho\left(\sum_{i=1}^{n\cdot 2^n-1} X_{i,1}^*; Z\right)= \rho(X;Z).
\end{align}

\vspace{0.2cm}

Write $ X_{n,1} := \sum_{i=1}^{n\cdot 2^n-1} X_{i,1}^*$.
For any positive integer $n \geq 1$, by (\ref{5014})--(\ref{5023}), we know that
\begin{align*}
 X_{n,1}(\omega)
   = \begin{cases}
           n - \frac{1}{2^n}, \quad\quad\quad & \mbox{if}\quad \tau_Z(X;Z(\omega))> n, \\
           X(\omega)-\frac{1}{2^n}, \quad\quad\quad & \mbox{if}\quad \frac{1}{2^n}< \tau_Z(X;Z(\omega))\leq n, \\
           0, \quad\quad\quad\quad\quad & \mbox{if}\quad 0 < \tau_Z(X;Z(\omega))\leq \frac{1}{2^n}, \\
           X(\omega), \quad\quad\quad\quad & \mbox{if}\quad \tau_Z(X;Z(\omega)) = 0,
    \end{cases}
\end{align*}
and that for any $z \in N^c \cap \mbox{Ran}(Z),$
\begin{align*}
 \tau_Z( X_{n,1}; z)
     = \begin{cases}
           n - \frac{1}{2^n}, \quad\quad\quad & \mbox{if}\quad \tau_Z(X; z)> n, \\
           \tau_Z(X; z),       \quad\quad\quad & \mbox{if}\quad \frac{1}{2^n}< \tau_Z(X; z)\leq n, \\
           0, \quad\quad\quad\quad\quad & \mbox{if}\quad 0 \leq \tau_Z(X; z)\leq \frac{1}{2^n}.
       \end{cases}
\end{align*}
Clearly, $ -\frac{1}{2} \leq X_{n, 1}$ for each $n \geq 1,$ and $\tau_Z( X_{n,1}; z) \leq \tau_Z( X; z) $ for each $n \geq 1$ and
each $z \in N^c \cap \mbox{Ran}(Z)$. Moreover, $X_{n,1} \uparrow X$ eventually.
Indeed, for any $\omega \in \Omega$, $\tau_Z(X;Z(\omega)) \geq 0$, since $X \in \sX_+$ and $\tau_Z(0; Z(\omega)) \leq \tau_Z(X;Z(\omega)).$
If $\tau_Z(X;Z(\omega)) = 0$, then for any $n \geq 1$, $X_{n,1}(\omega) = X(\omega)$.
If $\tau_Z(X;Z(\omega)) > 0$, then there exists some integer $N := N(\omega) \geq 1$ such that
$
\frac{1}{2^N} < \tau_Z(X; Z(\omega)) \leq N.
$
Hence, for all $n \geq N$,
\begin{align*}
    \frac{1}{2^n} \leq \frac{1}{2^N} < \tau_Z(X;Z(\omega)) \leq N \leq n,
\end{align*}
and thus $X_{n,1}(\omega) = X(\omega) - \frac{1}{2^n}$ for all $n \geq N$.
In summary, $X_{n,1} \uparrow X$ eventually. Therefore, by Axiom B4, we have that
\begin{align*}
\lim_{n\rightarrow \infty} \rho\left(\sum_{i=1}^{n\cdot 2^n-1} X_{i,1}^*; Z\right) = \rho(X;Z),
\end{align*}
which shows that (\ref{5007}) holds.

\vspace{0.2cm}

Next, we proceed to show that
\begin{align}\label{5006}
\lim_{n\rightarrow \infty} \rho\left( \sum_{i=1}^{n\cdot 2^n-1} X_{i,2}^*;Z \right) = \rho(X;Z).
\end{align}

\vspace{0.2cm}

Write $ X_{n,2} := \sum_{i=1}^{n\cdot 2^n-1} X_{i,2}^*$.
For any positive integer $ n \geq 1, $ by (\ref{5025})--(\ref{5028}), we have that
\begin{align*}
  X_{n,2}(\omega)
     = \begin{cases}
               n - \frac{1}{2^n}, \quad\quad\quad & \mbox{if}\quad \tau_Z(X; Z(\omega)) > n - \frac{1}{2^n}, \\
               X(\omega), \quad\quad\quad & \mbox{if}\quad 0 \leq \tau_Z(X; Z(\omega)) \leq n - \frac{1}{2^n},
      \end{cases}
\end{align*}
and that for any $z \in N^c \cap \mbox{Ran}(Z)$,
\begin{align*}
 \tau_Z( X_{n,2}; z)
     = \begin{cases}
               n - \frac{1}{2^n}, \quad\quad\quad & \mbox{if}\quad \tau_Z(X; z) > n - \frac{1}{2^n}, \\
               \tau_Z( X; z), \quad\quad\quad & \mbox{if}\quad 0 \leq \tau_Z(X; z) \leq n - \frac{1}{2^n}.
      \end{cases}
\end{align*}
Clearly, $0 \leq X_{n,2}$ for each $n \geq 1,$ $X_{n,2} \uparrow X$ eventually and  $\tau_Z( X_{n,2}; z) \leq \tau_Z( X; z) $ for each $n \geq 1$
and each $z \in N^c \cap \mbox{Ran}(Z)$. Hence, by Axiom B4, we know that
\begin{align*}
      \lim_{n\rightarrow \infty} \rho\left( \sum_{i=1}^{n\cdot 2^n-1} X_{i,2}^*;Z \right) = \rho(X;Z),
\end{align*}
which shows that (\ref{5006}) holds. Now, (\ref{5037}) follows from (\ref{5007})  and (\ref{5006}).

\vspace{0.2cm}

Step three. We  claim that
\begin{align}\label{5031}
 \lim_{n\rightarrow \infty}
   & \frac{1}{2^n}\sum_{i=1}^{n\cdot 2^n-1} \gamma_Z\left\{ z \in N^c \cap \mbox{Ran}(Z): \tau_Z(X;z)> \frac{i}{2^n} \right\} \nonumber \\
   & = \int_0^{\infty} \gamma_Z\{ z \in N^c \cap \mbox{Ran}(Z) : \tau_Z(X;z)>\alpha \} d\alpha.
\end{align}
In fact, since $\gamma_Z\{z \in N^c \cap \mbox{Ran}(Z) : \tau_Z(X;z)> s\}$ is decreasing with respect to variable $s$,
\begin{align*}
 \int_{\frac{1}{2^n}}^{n} & \gamma_Z \left\{z \in N^c \cap \mbox{Ran}(Z) : \tau_Z(X;z)> s \right\} ds \\
  & \leq  \frac{1}{2^n} \cdot \sum_{k=1}^{n\cdot 2^n - 1} \gamma_Z \left\{z \in N^c \cap \mbox{Ran}(Z) : \tau_Z(X;z)> \frac{k}{2^n} \right\}  \\
  & = \sum_{i=0}^{n\cdot 2^n - 2} \gamma_Z \left\{z \in N^c \cap \mbox{Ran}(Z) : \tau_Z(X;z)> \frac{i+1}{2^n} \right\} \cdot \frac{1}{2^n} \\
  & \leq \sum_{i=0}^{n\cdot 2^n - 2}  \int_{\frac{i}{2^n}}^{\frac{i+1}{2^n}} \gamma_Z\{z \in N^c \cap \mbox{Ran}(Z) : \tau_Z(X;z)> s\}ds \\
  & \leq \int_0^{\infty} \gamma_Z\{z \in N^c \cap \mbox{Ran}(Z) : \tau_Z(X;z)> s\} ds.
\end{align*}
Taking $n \rightarrow \infty$ in both sides of above inequality results in
\begin{align*}
\lim_{n\rightarrow \infty}
   & \frac{1}{2^n}\sum_{i=1}^{n\cdot 2^n-1}  \gamma_Z\left\{ z \in N^c \cap \mbox{Ran}(Z) : \tau_Z(X;z)> \frac{i}{2^n} \right\} \nonumber \\
   & = \int_0^{\infty} \gamma_Z\{z \in N^c \cap \mbox{Ran}(Z) : \tau_Z(X;z)> \beta\} d\beta,
\end{align*}
which is exactly (\ref{5031}).

\vspace{0.2cm}

As a conclusion, by (\ref{5018}), (\ref{5037}) and (\ref{5031}), we know that
$$
\rho(X; Z) =  \int_0^{\infty} \gamma_Z\{ z \in N^c \cap \mbox{Ran}(Z) : \tau_Z(X;z)>\alpha \} d\alpha,
$$
which, together with (\ref{5001}), yields that
$$
\rho(X;Z)=\int_0^{\infty} \gamma_Z\left\{ z \in N^c \cap \mbox{Ran}(Z) : \int_0^{\infty} g_z \circ K_Z(z,\{ X > \alpha \})d\alpha >\beta \right\}d\beta.
$$
Theorem 3.1 is proved.

\vspace{0.2cm}

\noindent{\bf Proof of Theorem 3.2.}

\vspace{0.2cm}

Let a random environment $Z \in \sX^\perp $ be arbitrarily given, and let the family $\{ \tau_Z(\cdot; z) ;$  $z \in \mbox{Ran}(Z) \}$
of functionals be as in Assumption A. Recalling ($\ref{0429add1}$), for every $z \in N_0^c, $
\begin{align}\label{add5006}
  K_Z(z,\{ Z \in B \}) = 1_{B}(z) \quad \mbox{for\ any} \ B \in \sB(\mathbb{R}),
\end{align}
where the $P_Z-$null set $N_0 \in \sB(\mathbb{R})$ is as in ($\ref{0429add1}$).
By Theorem 3.1, we know that there is a $P_Z-$null set $N \in \sB(\mathbb{R})$ with $N_0 \subseteq N,$
so that there exist a family $\{g_z; z \in N^c \cap\mbox{Ran}(Z)\}$ of
left-continuous distortion functions defined by (\ref{add5002}) and a monotone and normalized set function $\gamma_Z$
on $\sB(\mbox{Ran}(Z))$ depending on $Z$ such that for any $X \in \sX_+$,
\begin{align}\label{5011}
\rho(X;Z) = \int_0^{\infty} \gamma_Z\left\{ z \in N^c \cap \mbox{Ran}(Z) :  \tau_Z(X; z) >\beta \right\}d\beta,
\end{align}
where for every $ z \in N^c \cap \mbox{Ran}(Z),$  $\tau_Z(\cdot; z) $ is given by
\begin{align}\label{add5011}
 \tau_Z(Y; z) =  \int_0^{\infty} g_z \circ K_Z(z,\{ Y > \alpha \})d\alpha, \quad Y \in \sX_+,
\end{align}
and for $Y \in \sX,$
\begin{align}\label{1016add1}
\tau_Z(Y; z) = \int_{-\infty}^0 \left[ g_z \circ K_Z(z,\{ Y > \alpha \}) - 1 \right] d\alpha
                 + \int_0^{\infty} g_z \circ K_Z(z,\{ Y > \alpha \})d\alpha.
\end{align}

\vspace{0.2cm}

We first claim that for any $N^c \cap B^* := N^c \cap B \cap \mbox{Ran}(Z) \in \sB(\mbox{Ran}(Z))$ with some $B \in \sB(\mathbb{R})$,
\begin{align}\label{5012}
\gamma_Z(N^c \cap B^*) := \gamma_Z(N^c \cap B \cap \mbox{Ran}(Z)) = \rho(1_{B}(Z);Z).
\end{align}
In fact, from (\ref{add5006}) and (\ref{5011}), it follows that
\begin{align*}
 \rho(1_{B}(Z);Z)
  & = \int_0^{\infty} \gamma_Z\left\{ z \in N^c \cap  \mbox{Ran}(Z) : \int_0^1 g_z \circ K_Z(z,\{ Z \in B \})d\alpha >\beta \right\}d\beta \nonumber \\
  & = \int_0^{\infty} \gamma_Z\left\{ z \in N^c \cap \mbox{Ran}(Z) :  1_ B(z) > \beta \right\}d\beta \nonumber \\
  & = \gamma_Z(N^c \cap B \cap \mbox{Ran}(Z)),
\end{align*}
which shows that (\ref{5012}) holds.

\vspace{0.2cm}

To show the theorem, it is sufficient for us to show that there exists a function $h_Z : [0,1] \rightarrow [0, 1]$ with $h_Z(0) = 0$ and $h_Z(1) = 1$,
such that for any $t \geq 0$,
\begin{align}\label{5005}
  \gamma_Z\{ z \in N^c \cap \mbox{Ran}(Z) : \tau_Z(X;z) > t \} = h_Z \circ P_Z^* \{ z \in N^c \cap \mbox{Ran}(Z) : \tau_Z(X;z)>t \}.
\end{align}

\vspace{0.2cm}

For this purpose, write $\mbox{Ran}(P_Z^*) := \{ P_Z^*(N^c \cap B^*) : B^* \in \sB(\mbox{Ran}(Z)) \}$. Next, we proceed  to construct a function
$h_Z : \mbox{Ran}(P_Z^*) \rightarrow [0,1]$.
For every $u \in \mbox{Ran}(P_Z^*)$, there is some $B_u^* := B_u \cap \mbox{Ran}(Z) \in \sB(\mbox{Ran}(Z))$ with some $B_u \in \sB(\mathbb{R})$
such that $P_Z^*(N^c \cap B_u^*) = u$, and thus we define $h_Z$ as
\begin{align}\label{add5009}
h_Z(u) := \rho(1_{B_u}(Z);\ Z).
\end{align}

\vspace{0.2cm}

We first claim that $h_Z$ is well defined on $\mbox{Ran}(P_Z^*)$. In fact, for $u \in \mbox{Ran}(P_Z^*)$, let
$B_{u,1}^* :=  B_{u,1} \cap \mbox{Ran}(Z),\ B_{u,2}^* := B_{u,2} \cap \mbox{Ran}(Z) \in \sB(\mbox{Ran}(Z))$ with some
$B_{u,1}, B_{u,2} \in \sB(\mathbb{R})$ be such that $P_Z^*(N^c \cap B_{u,1}^*) = P_Z^*(N^c \cap B_{u,2}^*) = u$, then
$\tau_Z( 1_{B_{u,1}}(Z); \cdot)$ and $\tau_Z( 1_{B_{u,2}}(Z); \cdot)$ have the same probability distribution with respect to $P_Z^*$.
Indeed, for any $D \in \sB(\mathbb{R}),$  by (\ref{add5006}) and (\ref{add5011}),
\begin{align}\label{add5007}
  P_Z^* \circ \tau_Z^{-1}( 1_{B_{u,1}}(Z); \cdot) (D)
    & = P_Z^* \left\{ z \in N^c \cap \mbox{Ran}(Z) : \int_0^{\infty} g_z \circ K_Z(z,\{ 1_{B_{u,1}}(Z) > \alpha \})d\alpha  \in D \right\} \nonumber \\
    & = P_Z^* \left\{ z \in N^c \cap  \mbox{Ran}(Z) : \int_0^1 g_z \circ K_Z(z,\{ Z \in B_{u,1} \})d\alpha  \in D \right\} \nonumber \\
    & = P_Z^* \left\{ z \in N^c \cap  \mbox{Ran}(Z) :  1_{B_{u,1}}(z) \in D \right\} \nonumber \\
    & = P_Z^* \circ 1_{ B_{u,1}^* }^{-1} (D).
\end{align}
Similarly,
\begin{align}\label{add5008}
   P_Z^* \circ \tau_Z^{-1}( 1_{B_{u,2}}(Z); \cdot) (D) = P_Z^* \circ 1_{ B_{u,2}^* }^{-1} (D).
\end{align}
Note that
\begin{align*}
  P_Z^* ( 1_{ B_{u,1}^*} = 1 ) & =  P_Z^* ( B_{u,1}^*) = P_Z^* ( N^c \cap B_{u,1}^*) = u \\
                               & = P_Z^* (N^c \cap B_{u,2}^*) = P_Z^* ( B_{u,2}^*) = P_Z^* ( 1_{ B_{u,2}^*} = 1 ),
\end{align*}
which implies that $P_Z^* \circ 1_{ B_{u,1}^* }^{-1} = P_Z^* \circ 1_{ B_{u,2}^* }^{-1}$. Hence, keeping in mind (\ref{add5007}) and (\ref{add5008}),
we know that  $\tau_Z( 1_{B_{u,1}}(Z); \cdot)$ and $\tau_Z( 1_{B_{u,2}}(Z); \cdot)$ have the same probability distribution with respect to $P_Z^*$.
Therefore, by Axiom B1, we conclude that $\rho(1_{B_{u,1}}(Z); Z) = \rho(1_{B_{u,2}}(Z); Z)$, which means that
$h_Z$ defined by (\ref{add5009}) is well defined.

\vspace{0.2cm}

We further claim that $h_Z(0) = 0$ and $h_Z(1) =1.$ In fact, setting $B_0 := \emptyset$ implies $B^*_0 := B_0 \cap \mbox{Ran}(Z) = \emptyset.$
Thus $h_Z (0) =  \rho(1_\emptyset(Z); Z) =  \rho(0; Z) = 0,$ where the last equality is guaranteed by the positive homogeneity of $\rho(\cdot; Z)$
(see Remark 3.1(ii)). Similarly, setting $B_1 := \mathbb{R}$ implies $B^*_1 := B_1 \cap \mbox{Ran}(Z) = \mbox{Ran}(Z).$
Thus $h_Z (1) =  \rho(1_\mathbb{R}(Z); Z) =  \rho(1; Z) = 1,$  due to the normalization of $\rho$.
Moreover, it is clear that the range of $h_Z$ is contained in $[0, 1].$

\vspace{0.2cm}

Now, we can arbitrarily extend $h_Z$ from $\mbox{Ran}(P_Z^*)$ to the whole interval $[0,1]$, say by linear interpolation, because those $u \notin \mbox{Ran}(P_Z^*)$ do not matter. We still denote by $h_Z$ this extension, because it should have no risk of notation confusion.

\vspace{0.2cm}

Next, we proceed to show that (\ref{5005}) holds for $h_Z$ defined by (\ref{add5009}).
Given $X \in \sX_+$, for any $t\geq 0$, denote $B_t^*:= \{ z \in \mbox{Ran}(Z) : \tau_Z(X; z)>t \}$, then $B_t^* \in \sB(\mbox{Ran}(Z)),$
since $\{ \tau_Z(X; z); z\in \mbox{Ran}(Z) \}$
is regular on $\mbox{Ran}(Z)$. Hence, there exists some $B_t \in \sB(\mathbb{R})$ such that $ B_t^* = B_t \cap \mbox{Ran}(Z). $
From (\ref{5012}) and the definition of $h_Z$ as in (\ref{add5009}), it follows that
\begin{align*}
   h_Z \circ P_Z^*(N^c \cap B_t^*) :=  h_Z(P_Z^*(N^c \cap B_t^*)) = \rho(1_{B_t}(Z);Z) = \gamma_Z (N^c \cap B^*_t),
\end{align*}
That is,
\begin{align*}
  \gamma_Z\left\{ z \in  N^c \cap \mbox{Ran}(Z) : \tau_Z(X;z) > t \right\} = h_Z \circ P_Z^* \{ z \in N^c \cap \mbox{Ran}(Z) : \tau_Z(X;z)>t \}.
\end{align*}
which just shows that (\ref{5005}) holds for $h_Z$ defined by (\ref{add5009}). Theorem 3.2 is proved.

\vspace{0.2cm}

\noindent{\bf Proof of Theorem 3.3.}

\vspace{0.2cm}

The argumentation is basically the same as that of extending Proposition 3.3 from non-negative random variables to real-valued random variables.
We sketch the proof here.

\vspace{0.2cm}

Let a random environment $Z \in \sX^\perp $ be fixed. Given any $X \in \sX$,
then for any $m < 0$, $X \vee m - m \in \sX_{+}$. By Theorem 3.2 we know that
\begin{align}\label{huadd501}
    \rho(X \vee m - m; Z) = \int_0^{\infty} h_Z \circ P_Z^* \left\{ z \in N^c \cap \mbox{Ran}(Z) : \tau_Z(X \vee m - m; z) > \beta \right\} d\beta,
\end{align}
where the $P_Z-$null set $N$ and the function $h_Z$ are as in Theorem 3.2,
and for every $z \in N^c \cap \mbox{Ran}(Z),$ by change-of-variable,
\begin{align}\label{huadd502}
    \tau_Z(X \vee m - m; z) = \int_m^{\infty} g_z \circ K_Z(z, \{ X > \alpha \}) d\alpha,
\end{align}
where the distortion function $g_z$ is as in Theorem 3.2.

\vspace{0.2cm}

Observing the facts that the two functions $\tau_Z(X \vee m; \cdot)$ and $\tau_Z(-m; \cdot) = -m$ are apparently environment-wise comonotonic,
and that $X \vee m$ and $-m$ are local-comonotonic on $\{ Z=z \}$ for each $z \in \mbox{Ran}(Z)$,
by Axiom B3, we have that
\begin{align}\label{huadd503}
    \rho(X \vee m - m; Z) = \rho(X \vee m; Z) + \rho(-m; Z) = \rho(X \vee m; Z) + (-m),
\end{align}
where the fact that $\rho(-m; Z) = -m$ has been used.
Thus, from (\ref{huadd501})--(\ref{huadd503}) and change-of-variable, it follows that
\begin{align}\label{huadd504}
 \rho & (X  \vee m; Z) \nonumber \\
    & = \rho(X \vee m -m; Z) - (-m) \nonumber \\
    & = \int_m^{0} \left[ h_Z \circ P_Z^* \left\{ z \in N^c \cap \mbox{Ran}(Z) : \int_m^{\infty} g_z \circ K_Z(z, \{ X > \alpha \}) d\alpha + m
           > \beta \right\} -1 \right] d\beta \nonumber \\
    & \quad + \int_0^{\infty} h_Z \circ P_Z^* \left\{ z \in N^c \cap \mbox{Ran}(Z) : \int_m^{\infty} g_z \circ K_Z(z, \{ X > \alpha \}) d\alpha +m >\beta
                                                     \right\} d\beta.
\end{align}

\vspace{0.2cm}

Since $X$ is bounded,  we can choose $ m $ small enough, for example less than $ - \|X\|$, so that
\begin{align}\label{doublestar4}
 \rho(X \vee m; Z) =  \rho(X; Z)
\end{align}
and
\begin{align}\label{huadd505}
  \int_m^{\infty}
      & g_z \circ K_Z(z, \{ X > \alpha \}) d\alpha + m \nonumber \\
      & = \int_{-\infty}^0 \left[ g_z \circ K_Z(z, \{ X > \alpha \}) -1 \right] d\alpha + \int_0^{\infty} g_z \circ K_Z(z, \{ X > \alpha \}) d\alpha.
\end{align}
After plugging (\ref{doublestar4}) and (\ref{huadd505}) into (\ref{huadd504}), then letting $ m \rightarrow -\infty$ in both sides of
(\ref{huadd504}) results in the desired assertion $(\ref{3002})$. Theorem 3.3 is proved.

\vspace{0.2cm}

\noindent{\bf Proof of Theorem 3.4.}

\vspace{0.2cm}

Taking ($\ref{add3002}$) and ($\ref{add3012}$) into account, ($\ref{3001}$) can be rewritten as
\begin{align}\label{add5003}
  \rho_{_Z}(X)
     = \int \rho_{_Z}(X; \cdot) d h_Z \circ P_Z, \quad X \in \sX,
\end{align}
where $\rho_{_Z}(\cdot; z)$ is defined by ($\ref{add3001}$), that is, for every $z \in \mathbb{R}$,
\begin{align}\label{add5004}
  \rho_{_Z}(X; z) = \int X d g_z\circ K_Z(z,\cdot).
\end{align}
Therefore, the monotonicity is straightforward. By change-of-variable, elementary calculations can show the translation invariance
and positive homogeneity. Next, we simply check the subadditivity.
In fact, since the distortion functions $g_z$ and $h_Z$ are concave, by Example 2.1 of Denneberg (1994) or
Proposition 4.75 of F\"{o}llmer and Schied (2016), we know that for each $ z \in \mathbb{R}$,
$g_z \circ K_Z(z, \cdot)$ is a monotone and submodular set function on $\sF$, and that $h_Z \circ P_Z$  is a monotone and submodular
set function on $\sB(\mathbb{R})$. Hence, from (\ref{add5004}) and the Subadditivity Theorem of Denneberg (1994, Theorem 6.3),
it follows that for each $ z \in \mathbb{R}$ and $X_1, X_2 \in \sX$,
\begin{align*}
\rho_{_Z}(X_1 + X_2; z) \leq \int X_1  d g_z \circ K_Z(z, \cdot) + \int X_2  d g_z \circ K_Z(z, \cdot) = \rho_{_Z}(X_1; z) + \rho_{_Z}(X_2; z),
\end{align*}
which, together with (\ref{add5003}), the monotonicity of Choquet integral and the Subadditivity Theorem of Denneberg (1994, Theorem 6.3),
results in
\begin{align*}
\rho_{_Z}( X_1 + X_2) \leq \int \rho_{_Z}(X_1; \cdot) d h_Z \circ P_Z + \int \rho_{_Z}(X_2; \cdot) d h_Z \circ P_Z
                   = \rho_{_Z}( X_1) + \rho_{_Z}(X_2).
\end{align*}
Theorem 3.4 is proved.

\vspace{0.2cm}

\noindent{\bf Proof of Theorem 3.5.}

\vspace{0.2cm}

Keep (\ref{add5003}) and (\ref{add5004}) in mind.
Since $h_Z$ is concave, by Example 2.1 of Denneberg (1994) or Proposition 4.75 of F\"{o}llmer and Schied (2016),
we know that $h_Z \circ P_Z$ is a monotone, normalized and submodular set function on $\sB(\mathbb{R})$. Applying Theorem 4.94
of F\"{o}llmer and Schied (2016) to $h_Z \circ P_Z$ implies that
\begin{align}\label{add5012}
\int \rho_{_Z}(X; \cdot) d h_Z \circ P_Z =  \underset{Q \in \sQ_1}{\mbox{sup}} E_Q(\rho_{_Z}(X; \cdot)),
\end{align}
where $ \sQ_1 := \{Q_1 \in \sM_{1,f}(\mathbb{R}, \sB(\mathbb{R})) : Q_1(B) \leq h_Z \circ P_Z(B)\ \mbox{for\ all}\ B \in \sB(\mathbb{R})\},$
and the supremum taken over $\sQ_1$ can be attained at some $Q_{1, X} \in \sQ_1,$ that is,
\begin{align}\label{add5013}
\int \rho_{_Z}(X; \cdot) d h_Z \circ P_Z =  E_{Q_{1, X}}(\rho_{_Z}(X; \cdot)) = \int \rho_{_Z}(X; \cdot)  d Q_{1,X}.
\end{align}
Similarly, for each $z \in \mathbb{R}$, since $g_z$ is concave, then $g_z \circ K_Z(z, \cdot)$ is a monotone, normalized and submodular set function
on $ \sF.$  Again applying Theorem 4.94 of F\"{o}llmer and Schied (2016) to $g_z \circ K_Z(z, \cdot),$ we have that
\begin{align}\label{add5014}
\rho_{_Z}(X; z)  =  \underset{Q \in \sQ_{2, z}}{\mbox{sup}} E_Q(X),
\end{align}
where $\sQ_{2, z} := \{Q_{2,z} \in \sM_{1,f}(\Omega, \sF) :  Q_{2,z}(A) \leq g_z \circ K_Z(z, A)\ \mbox{for\ all}\ A \in \sF \},$
and the supremum taken over $\sQ_{2,z}$ can be attained at some $Q_{2, z, X} \in \sQ_{2, z},$ that is,
\begin{align}\label{add5015}
\rho_{_Z}(X; z) =  E_{Q_{2, z, X}}(X) = \int X  d Q_{2, z, X}.
\end{align}

\vspace{0.2cm}

By repeating integration  in
both sides of (\ref{add5015}) with respect to $\widehat{Q}_{1, X},$ the outer set function of $Q_{1, X}$, we have that
\begin{align*}
 \int \rho_{_Z}(X; z) \widehat{Q}_{1, X}(dz)
    = \int \rho_{_Z}(X; z) Q_{1, X}(dz)
    = \int \left( \int X(\omega) Q_{2, z, X}(d\omega) \right) Q_{1, X}(dz),
\end{align*}
which, together with (\ref{add5003}) and (\ref{add5013}), gives rise to
\begin{align}\label{add5018}
 \rho_{_Z}(X) = \int \left( \int X(\omega) Q_{2, z, X}(d\omega) \right) Q_{1, X}(dz).
\end{align}

\vspace{0.2cm}

On the other hand, given any $(Q_1, Q_2) \in \sC,$ where $Q_1 \in \sQ_1$, $ Q_2 := \{ Q_{2,z}\in\sQ_{2,z} ; z \in \mathbb{R} \},$
then for each $z \in \mathbb{\mathbb{R}}$,  both (\ref{add5014}) and  (\ref{add5015}) together yield
\begin{align}\label{add5019}
\int X(\omega) Q_{2, z}(d\omega)
   = E_{Q_{2, z}}(X)  \leq  \rho_{_Z}(X; z)  = \int X(\omega)  Q_{2, z, X}(d\omega).
\end{align}
Repeating integration in both sides of (\ref{add5019})  with respect to $\widehat{Q}_1,$ the outer set function of $Q_1$,
by (\ref{add5012}), (\ref{add5013}), (\ref{add5015}) and (\ref{add5019}), we have that
\begin{align}\label{add5020}
 \int \left( \int X(\omega) Q_{2, z}(d\omega) \right) \widehat{Q}_1(dz)
          \leq  \int  \left( \int X(\omega) Q_{2, z, X}(d\omega) \right) \widehat{Q}_{1,X}(dz).
\end{align}
Observing that $ (Q_{1,X}, Q_{2, z, X}) \in \sC,$ the desired assertion ($\ref{add3013}$) follows (\ref{add5018})
and (\ref{add5020}). Theorem 3.5 is proved.

\end{document}